\documentclass[11pt,a4paper]{article}
\pdfoutput=1
\usepackage[utf8]{inputenc}
\usepackage[T1]{fontenc}
\usepackage{lmodern}
\pdfmapfile{+lm.map}
\usepackage{amsmath,amssymb,mathtools}
\usepackage[margin=1in]{geometry}
\usepackage{microtype}
\usepackage{booktabs,longtable,array,calc}
\usepackage{caption}
\usepackage{float}
\usepackage{graphicx}
\usepackage{xcolor}
\usepackage{listings}
\usepackage{newunicodechar}
\usepackage{hyperref}
\usepackage{bookmark}
\usepackage{xurl}
\hypersetup{colorlinks=true,linkcolor=blue!55!black,urlcolor=blue!55!black,pdftitle={Coexponential Exploration in the Clef Programming Language: A Fearless Look Into How the Fidelity Framework Weaves Stateful Services and Delimited Continuations Across Multiple Hardware Targets},pdfauthor={Houston Haynes}}
\newcommand{\pandocbounded}[1]{\begingroup\setbox0\hbox{#1}\ifdim\wd0>\linewidth\resizebox{\linewidth}{!}{#1}\else#1\fi\endgroup}
\newsavebox{\paperdisplaybox}
\let\paperdisplayopen\[
\let\paperdisplayclose\]
\long\def\[#1\]{\begingroup
  \sbox{\paperdisplaybox}{$\displaystyle #1$}%
  \ifdim\wd\paperdisplaybox>1.25\linewidth
    \PackageError{paper}{Display requires source line breaks}{Split the display in Markdown.}%
  \fi
  \paperdisplayopen
  \ifdim\wd\paperdisplaybox>\linewidth
    \typeout{Paper display scaled to fit width: \the\wd\paperdisplaybox}%
    \resizebox{\linewidth}{!}{\usebox{\paperdisplaybox}}%
  \else\usebox{\paperdisplaybox}\fi
  \paperdisplayclose\endgroup}
\newunicodechar{α}{\ensuremath{\alpha}}
\newunicodechar{β}{\ensuremath{\beta}}
\newunicodechar{ε}{\ensuremath{\varepsilon}}
\newunicodechar{θ}{\ensuremath{\theta}}
\newunicodechar{κ}{\ensuremath{\kappa}}
\newunicodechar{Σ}{\ensuremath{\Sigma}}
\newunicodechar{⊗}{\ensuremath{\otimes}}
\newunicodechar{⊕}{\ensuremath{\oplus}}
\newunicodechar{↔}{\ensuremath{\leftrightarrow}}
\newunicodechar{¬}{\ensuremath{\neg}}
\newunicodechar{∧}{\ensuremath{\land}}
\newunicodechar{≤}{\ensuremath{\leq}}
\newunicodechar{×}{\ensuremath{\times}}
\newunicodechar{·}{\ensuremath{\cdot}}
\newunicodechar{†}{\ensuremath{\dagger}}
\newunicodechar{⟩}{\ensuremath{\rangle}}
\newunicodechar{²}{\ensuremath{{}^2}}
\newunicodechar{¹}{\ensuremath{{}^1}}
\newunicodechar{⁻}{\ensuremath{{}^-}}
\newunicodechar{ᵢ}{\ensuremath{{}_i}}
\newunicodechar{₀}{\ensuremath{{}_0}}
\newunicodechar{₁}{\ensuremath{{}_1}}
\newunicodechar{₂}{\ensuremath{{}_2}}
\newunicodechar{₊}{\ensuremath{{}_+}}
\begin{document}
\begin{center}

\textbf{Coexponential Exploration in the Clef Programming Language}

\emph{A Fearless Look Into How the Fidelity Framework Weaves Stateful Services and Delimited Continuations Across Multiple Hardware Targets}

\textbf{Houston Haynes}\\
Braidpoint, Asheville, NC\\
\href{mailto:research@braidpoint.tech}{\nolinkurl{research@braidpoint.tech}}\\
October 2026

\end{center}

\section*{Abstract}\label{abstract}
\addcontentsline{toc}{section}{Abstract}

Clef was created as a concurrent programming language whose compiler tracks state ownership, waiting computations and request dependencies. We develop a coexponential account of stateful services and their client pools within this design. Delimited continuations retain the computation and resources needed to complete each interaction. These structures give our compiler premises for verifying service composition and preserving it through native lowering to multiple hardware targets.

We connect finite actor state transitions to the initial, service and final rules of a coexponential server. A separate operational account relates actor dispatch to single-use continuation frames. We prove state-ownership and reply preservation, give a finite communication-summary analysis, and derive circular-wait exclusion under stated assumptions. A further result identifies when changing service order preserves both state and replies.

Our Program Semantic Graph retains these facts through compilation. We are researching reusable proofs to confirm the ownership and communication interfaces, while our BAREWire implementation connects message meaning to its unboxed representation across respective endpoints. A distributed worker example preserves request identity across suspension, reordered replies and boundary conversion. Urschel\textquotesingle s growth bounds inform a solver service; a distributed ADM update additionally preserves joint samples, metric evidence and request identity.

Our two-category account connects component refinement, reusable proof terms and native realization. Protocols determine the admissible interactions, continuations expose their storage requirements, and retained dependencies support differential compilation. Developers can express ordinary stateful services while the compiler carries the conditions for their correct composition across execution targets.

\section*{1. Concurrent Composition in Clef}\label{concurrent-composition-in-clef}
\addcontentsline{toc}{section}{1. Concurrent Composition in Clef}

We design Clef around computations that can proceed independently, exchange values and wait reliably on explicitly represented dependencies. Our compiler retains the waiting invocation, the state and resources it owns, and the dependencies required for its resumption. These relationships give stateful services a common basis for checking ownership, communication and completion.

Consider a service that accepts a request, changes its private state and returns a result. Its input and output types can be correct even when the result reaches the wrong caller. The state can remain free of data races while two services wait indefinitely for one another. A timeout can release a caller while the remote request continues to change state. Each failure concerns a relationship between otherwise well-typed operations.

An actor makes several of these relationships visible. It owns state, receives messages through an interface and selects work according to an execution policy. A request that waits for another service also identifies a suspended computation. The values retained by that computation, its right to resume and the work it awaits belong to the same program structure. Our aim is to make that structure available to the compiler, together with the proofs that justify its use.

The practical lineage is familiar. Erlang developed process isolation, message passing and supervision into an engineering discipline for long-running services. Armstrong\textquotesingle s account dates its development to 1986 \hyperref[ref-armstrong]{{[}1{]}}. In our Clef design, we connect those concerns to native compilation and automatic verification. A developer can use a stateful service through an ordinary library interface while the compiler retains the resource, protocol and dependency information needed to check its composition.

We organize the account around three questions. A message\textquotesingle s data contract describes its values and representation. A session protocol describes the permitted conversation. A wait relation describes the work that must complete before another computation can continue. Our site\textquotesingle s \emph{Three-Layer Actor Contract} introduces this division. Here we give it a formal construction and connect it to the control operations that execute the conversation.

\subsection*{1.1 Stateful Service as a Type-Theoretic Construction}\label{stateful-service-as-a-type-theoretic-construction}
\addcontentsline{toc}{subsection}{1.1 Stateful Service as a Type-Theoretic Construction}

Our starting point is a service used by several independent clients whose requests share one evolving state. We use the coexponentials introduced by Qian, Kavvos and Birkedal to express this arrangement in classical linear logic \hyperref[ref-qkb]{{[}2{]}}. Their client-server calculus supplies rules for forming a client pool and threading state through its service. The contribution of this paper is the connection between those rules, a stated actor fragment and the continuation representation retained by our compiler.

Here \emph{coexponential} names a unary modality over a session protocol. A separate use of the term concerns categorical duals of function types, given computational interpretations by Choudhury and Gay \hyperref[ref-cofunction]{{[}3{]}}. In our construction, the modalities describe a shared service and its client pool. Delimited continuations retain the control and resources of a suspended interaction within that arrangement.

We begin with a finite construction whose value encodings and handler derivations can be written explicitly. We then give an operational translation with local ownership and trace-preservation results. The compiler account extends these results through admitted value representations, finite communication analysis, library proofs and target contracts. This separates the parts supplied by a typing derivation from the additional properties of scheduling, arithmetic and network execution.

The distinction has a direct benefit. A proof that state ownership survives suspension can serve many handlers. A proof about a reliable channel can serve many conversations. A proof that a particular state update commutes with another can permit an additional scheduling choice. Our compiler obtains the participating values and operations from the program and connects the applicable proof terms to those sites.

\subsection*{1.2 Retaining the Relationships Through Compilation}\label{retaining-the-relationships-through-compilation}
\addcontentsline{toc}{subsection}{1.2 Retaining the Relationships Through Compilation}

Our Program Semantic Graph (PSG) retains source identities, control, resource use and communication relationships. These meet when independent work rejoins a continuation and its results determine subsequent work. The graph\textquotesingle s hyperedges represent obligations that depend jointly on several participants. A reply, for example, relates an invocation identity, an expected result type, a suspended continuation and its storage lifetime. Keeping those participants together lets a later transformation establish that it still realizes the same interaction.

Our companion papers supply the surrounding compiler account. \emph{Dimensional Type Systems and Deterministic Memory Management} develops dimensional meaning and memory lifetime \hyperref[ref-dts]{{[}4{]}}. \emph{The Program Hypergraph} develops joint relational structure \hyperref[ref-phg]{{[}5{]}}. Together, they describe the values, storage and relationships on which a communication contract depends.

Our \emph{Decidable by Construction} paper connects structural inference to local checks, reusable system proofs and relational reasoning \hyperref[ref-dbc]{{[}6{]}}. \emph{Fixed-Point Scaffolding} explains how those facts remain bound to operations through a thin middle end built on MLIR (Multi-Level Intermediate Representation) and through differential compilation \hyperref[ref-fps]{{[}7{]}}. The present paper supplies a stateful communication construction for those mechanisms to preserve.

Our site\textquotesingle s \emph{A Triangle Without Mystery} describes the common engineering discipline behind this work. Dimensional types retain a quantity\textquotesingle s meaning, checked encodings relate its representations, and compatible evidence connects the stages of compilation. A concurrent request brings these concerns together. Its value, invocation identity and resource authority must remain associated as it becomes a message, a suspended frame and a native operation. The categorical account developed below makes both kinds of composition explicit: connecting computations and composing justified changes to their realization. The compiler can then use the same relationships to preserve a service and to determine which established work survives a change to it.

\section*{2. A Service and Its Clients}\label{a-service-and-its-clients}
\addcontentsline{toc}{section}{2. A Service and Its Clients}

An ordinary function describes one application to arguments. A stateful service also connects that application to earlier and later requests. Reserving one of the last available resources changes what the next caller can receive. The service interface therefore needs to account for repeated interactions and their shared state.

\begingroup\floatplacement{figure}{htbp}\setkeys{Gin}{keepaspectratio}\captionsetup{labelformat=empty,font={small,sf},margin=1em,justification=raggedright,singlelinecheck=false}

\begin{figure}
\centering
\includegraphics[width=0.55\linewidth,height=0.4\textheight]{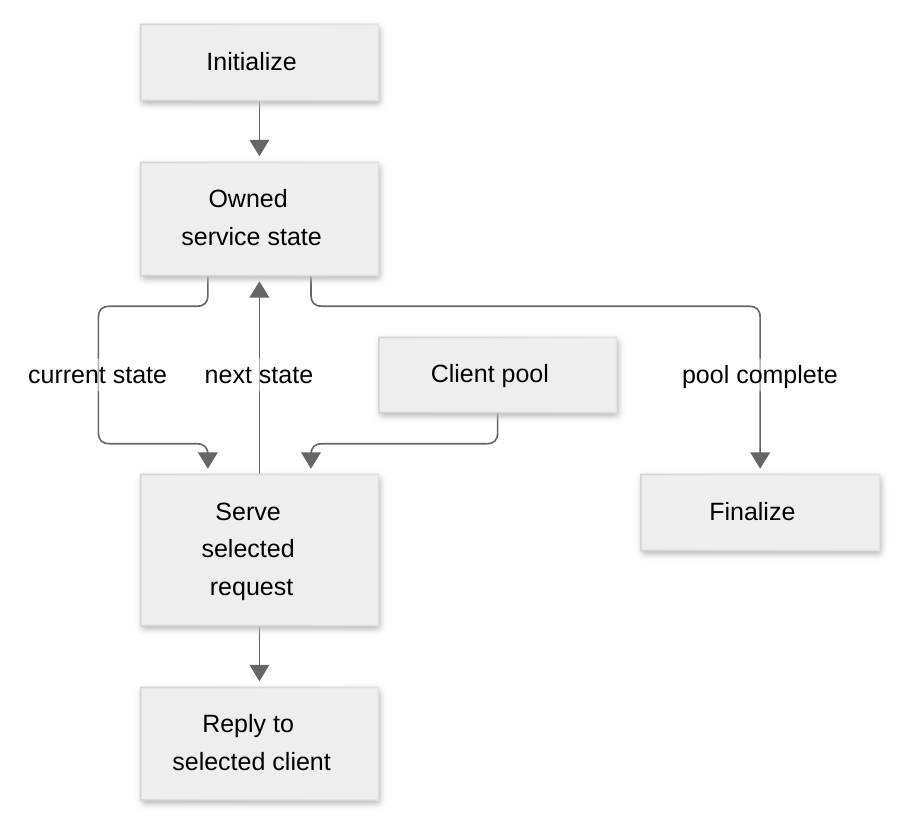}
\caption{\textbf{Figure 1.} A client pool shares one stateful service. Each selected handler transforms the private state and replies to its client. State arrows track exclusive ownership; finalization consumes the state when the pool is complete.}
\end{figure}

\endgroup

We write \(\mathsf{Srv}(A)\) for the server coexponential and \(\mathsf{Pool}(A)\) for its client-side counterpart. These are the modalities written ¡ and ¿ in Qian, Kavvos and Birkedal\textquotesingle s paper. Here \(A\) always denotes the server\textquotesingle s view of one interaction. Its clients use the dual protocol \(A^\perp\), so the connected pool has type \(\mathsf{Pool}(A^\perp)\).

\subsection*{2.1 The List Presentation}\label{the-list-presentation}
\addcontentsline{toc}{subsection}{2.1 The List Presentation}

For the operational construction we use the list presentation from §3.3 of the client-server sessions paper \hyperref[ref-qkb]{{[}2{]}}:

\[
\begin{aligned}
\mathsf{Srv}(A)&=\nu X.\bigl(\bot\mathbin{\&}(A\mathbin{\mathrm{par}}X)\bigr),\\
\mathsf{Pool}(A)&=\mu X.\bigl(1\oplus(A\otimes X)\bigr).
\end{aligned}
\]

The pool can be empty or contain a client followed by the rest of the pool. The server can finalize or provide an interaction while continuing with its remaining service. The symbols \(\mu\) and \(\nu\) denote least and greatest fixed points. The connectives \(\otimes\) (tensor) and \(\mathbin{\mathrm{par}}\) (par) are dual multiplicatives, and \(\oplus\) and \(\&\) describe the two directions of choice.

Our closing and empty-pool constructions use the M-True and QueW rules specified by Qian, Kavvos and Birkedal \hyperref[ref-qkb-errata]{{[}8{]}}. Both require the continuation premise \(P\vdash\varnothing\), an empty hypersequent. Their base processes are \lstinline!x[].stop! for closing and \lstinline!¿x[].stop! for the empty pool. Independent work is composed in parallel with these processes. This makes the independence recorded by the typing judgment explicit in the process structure.

The client rules form an empty pool and add a client from a separate component. This retains the client\textquotesingle s other interfaces. The server rule has three premises:

\[
\frac{
  \vdash\Gamma,B
  \qquad
  \vdash B^\perp,\Delta
  \qquad
  \vdash B^\perp,A,B
}{
  \vdash\Gamma,\Delta,\mathsf{Srv}(A)
}.
\]

Initialization provides the internal state protocol \(B\). Finalization consumes it. A worker consumes the previous state, serves one client through \(A\) and provides the next state. The external service type abstracts over this internal protocol. Replies and finalization can still export values specified by \(A\) and \(\Delta\).

This rule gives a precise account of a common actor arrangement. It also identifies the construction that an actor translation must supply. A declared value type for the state determines what data is retained. A session representation of that data determines how the state passes between workers.

\subsection*{2.2 Service Order and Observable Results}\label{service-order-and-observable-results}
\addcontentsline{toc}{subsection}{2.2 Service Order and Observable Results}

The pool\textquotesingle s construction permits service order to vary. In §3.4, Qian, Kavvos and Birkedal quotient pool formation by permutations so that reduction can select different clients \hyperref[ref-qkb]{{[}2{]}}. Their compare-and-set example deliberately uses the resulting race. A valid session can therefore return different results under different service orders.

Our actor interpretation preserves the runtime\textquotesingle s eligible-message policy. A first-in, first-out policy can select a restricted set of the orders admitted by the pool. A different policy can expose more of them. Each selection must still execute the worker protocol and transfer state correctly. An application that requires the same answers under several orders adds the condition developed in §6.

The fixed-point presentation describes the logical service interface. A concrete request batch has finite client construction. An indefinitely available service additionally needs admission, resource and scheduling contracts for its successive interactions. Those contracts determine how the persistent interface is realized within a bounded machine.

\section*{3. Constructing an Actor Service}\label{constructing-an-actor-service}
\addcontentsline{toc}{section}{3. Constructing an Actor Service}

The connection becomes explicit when state and messages have finite carriers. This case lets us derive the worker protocol directly and distinguish the data supplied by an actor declaration from the communication proof built around it. We generalize the Boolean compare-and-set construction in §4 of Qian, Kavvos and Birkedal\textquotesingle s paper to finite state, payload and reply sets and a family of message tags \hyperref[ref-qkb]{{[}2{]}}.

Let \(S\) be a finite nonempty state set. For each message tag \(m\) in a finite nonempty set \(M\), let \(P_m\) and \(R_m\) be finite nonempty payload and reply sets. A handler is a total transition

\[
h_m:S\times P_m\longrightarrow R_m\times S.
\]

The actor starts in a specified state \(s_0\). We use a nonreentrant request discipline. One handler owns the actor state until it returns its successor state, including any interval in which that handler is suspended. This provides a simple service order against which the translation can be checked.

\subsection*{3.1 From Values to an Internal Protocol}\label{from-values-to-an-internal-protocol}
\addcontentsline{toc}{subsection}{3.1 From Values to an Internal Protocol}

For a finite carrier \(X\), define

\[
E(X)=\bigoplus_{x\in X}1.
\]

A provider of \(E(X)\) selects the label for its value and closes its endpoint with \lstinline!c[].stop!. Its dual consumer receives that label and consumes the closing message. Finite sums can be represented through binary choices. Set

\[
\begin{aligned}
B&=E(S),\\
A&=\mathop{\&}_{m\in M}
\left(E(P_m)^\perp\mathbin{\mathrm{par}}
      \left(E(R_m)\otimes\bot\right)\right).
\end{aligned}
\]

From the server\textquotesingle s perspective, \(A\) offers the message tags, receives a payload-provider endpoint, returns a reply-provider endpoint and receives the end of the interaction. The payload and reply values travel through their respective encodings. The client follows the dual sequence.

\textbf{Proposition 1 (finite actor service).} Each family of total transitions \(h_m\) admits a worker derivation

\[
Q\vdash z:B^\perp,\ y:A,\ z':B
\]

that returns the reply and successor state specified by the selected transition. Initialization at \(s_0\) and a typed finalizer complete the premises of a coexponential server. For every selected order of a finite pool of independent, terminating clients following \(A^\perp\), the server has an execution with the same state sequence and labelled replies as the actor in that order.

\emph{Proof.} Offer each tag using \(\&\), receive its payload endpoint using \(\mathbin{\mathrm{par}}\), and branch on the state and payload encodings. In branch \((m,s,p)\), the table \(h_m\) fixes a pair \((r,s')\). Selecting the corresponding labels gives providers of \(E(R_m)\) and \(B\). Combine the reply provider with the remaining client close and the successor-state provider using \(\otimes\). Reassembling the exclusive branches gives the displayed worker judgment. Each branch uses the previous state once. Initialization selects the label \(s_0\), and a finalizer consumes \(B\) under its declared output protocol. The server rule then supplies \(\mathsf{Srv}(A)\). The pool starts with \lstinline!¿y[].stop!. Independent client processes are composed in parallel and added by the QueA rule. For a chosen client order, the pool permutation makes the next client available, and one worker computes the corresponding table entry. Induction over the pool size gives the same state sequence and request-indexed replies. The empty-pool interaction invokes finalization. \(\square\)

This instance of the list-form server and pool rules establishes a finite correspondence with a concrete value interpretation. The state declaration supplies \(S\), the encoding supplies \(B\), and the handler realization supplies the worker premise.

\subsection*{3.2 A Practical Representation Interface}\label{a-practical-representation-interface}
\addcontentsline{toc}{subsection}{3.2 A Practical Representation Interface}

The finite encoding is a mathematical construction. Compiling a machine word by enumerating every possible label would defeat its purpose. Our compiler instead admits value representations through typed provider and consumer operations with established communication and value-preservation laws. Word, record and buffer representations can then reuse those laws.

For a handler using such a representation, the worker obligation combines its state transition with its communication and resource use. Arithmetic and foreign calls carry their own contracts. State containing another service endpoint also retains that endpoint\textquotesingle s protocol and ownership. A pointer-sized representation gives the storage shape. The protocol supplies the permitted interactions.

This is where reusable library proofs remove repeated work. Once a value representation and a handler combinator have established laws, the compiler checks their actual parameters and participants. The developer writes the state type, message cases and computation. The compiler constructs the corresponding obligations from that structure.

\subsection*{3.3 Parameters, State and Compositional Structure}\label{parameters-state-and-compositional-structure}
\addcontentsline{toc}{subsection}{3.3 Parameters, State and Compositional Structure}

A handler can also depend on configuration shared across requests. A solver may retain a matrix, and several workers may use the same immutable parameter version. That relationship matters when the compiler specializes the handler or chooses where its data resides. We find a precise account of such composition in the parameterized construction used by Gavranović and colleagues \hyperref[ref-cdl]{{[}9{]}}.

Let \(\Theta\) be a configuration space. A parameterized handler has the form

\[
\mathsf{cell}_m:\Theta\times S\times P_m
\longrightarrow R_m\times S.
\]

Fixing \(\theta\in\Theta\) gives the transition \(h_m\) of §3. This is the shape of the parameterized Mealy cell in Example I.4 of their paper, where an input produces an output and a successor state. Our coexponential construction supplies the client pool and service protocol around that transition. The state ownership and invocation rules determine how concurrent clients may use it.

In the Cartesian instance of \(\operatorname{Para}(\mathcal C)\), data spaces are objects. A 1-cell from \(X\) to \(Y\) consists of a parameter space \(\Theta\) and a computation \(f:\Theta\times X\to Y\). A 2-cell compares such computations through a compatible reparameterization. For \(r:\Theta'\to\Theta\), the comparison from \((\Theta,f)\) to \((\Theta',f')\) requires

\[
f'(\theta',x)=f(r(\theta'),x).
\]

This equation gives the second level of structure a concrete purpose. A diagonal map \(\theta\mapsto(\theta,\theta)\) can express that two component uses draw on one configuration. The compiler must preserve that shared origin when it selects storage and update behavior. Our interpretation keeps evolving actor state under its ownership contract. A training operation that changes parameters includes that change in its state transition or in a separately specified update operation.

The parameterized account supplies algebra for computations and their relationships. Our compiler additionally needs comparisons between different operational realizations of the same service, including their effects, waits and resources. Section 10 gives that comparison its own two-category construction. A library can establish a permitted transformation once, and the compiler can apply it where the actual participants satisfy its premises. The application developer continues to work with the handler\textquotesingle s data and operations.

For an adaptive worker, one invocation can bind an immutable model configuration while the service retains evolving sampling or optimizer state. The request in §9.3 also fixes its parameter point, selected directions and geometric observation. A foreign sampler\textquotesingle s hidden state or callback environment belongs to that invocation\textquotesingle s effect and probability contract. Equal explicit arguments alone do not establish that two calls supply interchangeable observations.

The next step is to preserve the state and selected configuration while a handler waits. Its continuation must retain the particular interaction to which those values belong.

\section*{4. Executing the Protocol with Delimited Continuations}\label{executing-the-protocol-with-delimited-continuations}
\addcontentsline{toc}{section}{4. Executing the Protocol with Delimited Continuations}

A handler that calls another service may have work left after the reply. A delimited continuation represents that bounded remainder of the computation. It includes the values needed after suspension and a return point within an identified control scope. We use the delimited-control tradition of Danvy and Filinski \hyperref[ref-danvy-filinski]{{[}10{]}} to give this remainder an explicit operational role.

Our translation retains actor state inside the computation that owns it. Suspending a handler moves its live frame into a continuation. A matching completion resumes that frame. The coexponential describes the persistent service and its clients, while the continuation realizes the control needed within a particular interaction.

For a remote call, the caller retains that frame and its backing storage. The request carries its payload and a protocol-defined reply endpoint. The matching returned value resumes the waiting computation through that endpoint\textquotesingle s local binding. Section 7 supplies the resource-transfer and boundary conditions connecting the two sides.

\subsection*{4.1 The Retained Frame}\label{the-retained-frame}
\addcontentsline{toc}{subsection}{4.1 The Retained Frame}

For each invocation, let \(q\) be a fresh identity within an actor generation. A pending request carries its message tag, payload and a right to complete that invocation. A frame has the form

\[
\kappa=\langle q,\ell,\rho,\Omega\rangle,
\]

where \(\ell\) identifies the return point, \(\rho\) supplies the live values and \(\Omega\) records the resources owned by the suspended computation. In the fragment of §3, \(\Omega\) includes the actor\textquotesingle s exclusive state capability.

Our abstract actor execution uses the following transitions. The continuation realization implements the corresponding transfers.

{\def\LTcaptype{none} % do not increment counter
\begin{longtable}[]{@{}
  >{\raggedright\arraybackslash}p{(\linewidth - 4\tabcolsep) * \real{0.3333}}
  >{\raggedright\arraybackslash}p{(\linewidth - 4\tabcolsep) * \real{0.3333}}
  >{\raggedright\arraybackslash}p{(\linewidth - 4\tabcolsep) * \real{0.3333}}@{}}
\toprule\noalign{}
\begin{minipage}[b]{\linewidth}\raggedright
Actor transition
\end{minipage} & \begin{minipage}[b]{\linewidth}\raggedright
Continuation realization
\end{minipage} & \begin{minipage}[b]{\linewidth}\raggedright
Retained condition
\end{minipage} \\
\midrule\noalign{}
\endhead
\bottomrule\noalign{}
\endlastfoot
Dispatch a selected request & Enter the handler\textquotesingle s delimited computation & The same request and exclusive state are selected \\
Evaluate a local operation & Execute its admitted realization & Values and effects satisfy the operation\textquotesingle s contract \\
Await a dependent result & Capture the live frame & State, resources and the return point remain associated \\
Receive the matching completion & Consume the pending resumption right and restore the frame & One completion resumes the intended invocation \\
Return a reply and successor state & Complete the invocation and restore the actor\textquotesingle s idle state cell & The reply identity and state transition agree \\
\end{longtable}
}

The table defines a transfer discipline, including suspension within a handler. Further mailbox dispatch for this actor waits until the state returns. Other actors remain eligible to execute. Libraries that permit additional same-actor interleaving need the corresponding state-sharing invariant.

\subsection*{4.2 Local Preservation}\label{local-preservation}
\addcontentsline{toc}{subsection}{4.2 Local Preservation}

We relate a source configuration to its continuation realization by equality of mailbox entries, invocation identities, logical state values and pending or consumed completion rights. An idle state cell corresponds to an idle actor. A running or suspended handler corresponds to the frame that owns its state. Local operations have matching value and effect interpretations. Both executions use the same eligible-message policy.

Administrative target steps may allocate or select a frame, but preserve this relation and finish in finitely many steps. Capture and resumption preserve the frame\textquotesingle s return point, live values and owned resources. Completion consumes the invocation\textquotesingle s right exactly once. The correspondence in the table is a two-way step contract. Each enabled source transition has a finite matching target path, and each visible target transition matches an allowed source step. These requirements are contracts on the transition primitives used by the translation.

\textbf{Proposition 2 (conditional state and reply preservation).} Under these primitive contracts, the translation preserves unique actor-state ownership and at-most-once completion. Each finite source execution has a target execution with the same visible request, reply and state-transition events. Conversely, each finite target execution built from the stated transitions projects to such a source execution after administrative steps are removed.

\emph{Proof.} Initially the state belongs to one idle actor cell and each admitted invocation has one completion right. Dispatch transfers state to its handler. Local operations preserve the admitted ownership relation. Capture transfers the handler frame to one suspended continuation, and resumption consumes that continuation while restoring the same frame. Completion consumes one invocation right and transfers the successor state to the idle cell. None of these rules duplicates either authority. Induction preserves uniqueness and at-most-once completion. For the execution correspondence, dispatch, local evaluation, await, resumption and completion each have the matching source transition specified above. Administrative steps preserve the relation. Induction over either finite execution gives the corresponding sequence on the other side. \(\square\)

Our actor-to-continuation correspondence uses standard invariant and finite-trace arguments to give the compiler a local preservation contract. It names the resource and identity facts that each realization must maintain. The compiler can therefore admit single-use capture and resumption while retaining one owner of actor state and one authority for each completion.

\subsection*{4.3 Completion, Failure and Storage}\label{completion-failure-and-storage}
\addcontentsline{toc}{subsection}{4.3 Completion, Failure and Storage}

Our invocation interface can return a tagged outcome \(\mathsf{Reply}(r)\) or \(\mathsf{Failure}(e)\). A race between reply and timeout requires one atomic selection of the terminal outcome. The selected outcome resumes the pending frame once and enters its ordinary continuation or its cleanup path. Late messages retain their old invocation and generation identities, allowing the receiver to reject them after that right has been consumed.

This establishes the safety side of one-shot completion. Eventual resumption also requires the appropriate progress premises. A timeout means that the local wait has ended under its contract. The remote operation\textquotesingle s effects and cleanup are governed by the remote protocol. Operations that may be retried need a stated rule for request identity and repeated effects.

The captured frame also has a lifetime. A continuation used within a bounded local scope may use stack storage. A suspended request can require a region that survives several scheduler turns. Our PSG records the values live across each suspension cut and the lifetime of their backing storage. This lets frame layout and placement follow the computation\textquotesingle s needs while the same ownership proof remains attached.

Our use of single-use continuations has established implementation precedents. The effect-handler work of Sivaramakrishnan and colleagues uses one-shot continuations for concurrency in OCaml \hyperref[ref-ocaml-effects]{{[}11{]}}. Our contribution here is the connection to actor state, protocol evidence and the preservation obligations carried through native lowering.

\subsection*{4.4 Captures, Demand and Delivery}\label{captures-demand-and-delivery}
\addcontentsline{toc}{subsection}{4.4 Captures, Demand and Delivery}

Our flat closure representation gives the compiler a common structure for execution and proof dispatch. A callable associates its code with one environment containing its captures. A thunk retains work until its result is demanded. A delimited continuation retains the remainder of an active computation across a wait. Clef\textquotesingle s default lazy evaluation shares deferred computations by identity, including when they are passed or captured.

Our PSG associates these operations and their captured values with types, effects, lifetimes and resource permissions. The compiler uses this structure to select applicable laws, instantiate reusable proof terms and establish their premises for the actual computation. Laws governing demand, sharing and resumption can therefore justify optimizations and resource transfers at each admitted use. Their proof conclusions remain associated with the operations whose realization the compiler must preserve through lowering.

Eager evaluation selects where demand occurs. An explicit \lstinline!eager! expression demands its outer value at an active boundary, preserving the specified order of eager operands. Captured computations and nested lazy values retain their own demand conditions. For a shared lazy result, the library contract accounts for evaluation and waiting demands, including the permissions with which they receive the result. The compiler admits early evaluation or thunk elimination when its preservation argument covers returned values, effects, sharing, storage identity, lifetime and termination.

An observable supplies events to a subscribed observer. The subscription retains that observer\textquotesingle s captured environment for its admitted lifetime, including pending invocations. Activation and delivery remain separate choices. A cold producer can defer connection until demand and then deliver events by push. Our basic observable contract preserves synchronous delivery in subscription order. Buffered or asynchronous delivery adds its own ordering, capacity and cancellation rules. A subscription can invoke its observer many times, while each suspended invocation retains its own single-use completion right. The graph connects both lifetimes to the values and state they use.

Runtime incremental computation adds cached results, active dependencies and invalidation. When a stale result is demanded, its computation validates dependencies in dependency order and recomputes where required. Reuse accounts for actual reads and effects, including those reached through helper calls or aliases. A cutoff can stop propagation from an unchanged output while preserving other outstanding invalidations. Together, these structures expose when work is needed, what it may share, which event can resume it and which established result remains usable. The ownership and communication proofs can then be applied to the operations and resources that each choice actually uses.

\section*{5. Recovering Communication Structure}\label{recovering-communication-structure}
\addcontentsline{toc}{section}{5. Recovering Communication Structure}

Much of the required information already appears in an ordinary handler. Its message cases identify available requests. Typed calls identify payloads and results. An await identifies a dependency and a suspension cut. The compiler can recover this information without asking a developer to write a second description of the same control flow.

The recovery problem has several levels. A finite communication summary records possible actions. A session derivation additionally establishes how those actions compose with their peers and use their resources. A functional proof can establish what result the conversation computes. Our analysis retains each established fact with the operation and assumptions that support it.

\subsection*{5.1 A Finite Summary with a Stated Guarantee}\label{a-finite-summary-with-a-stated-guarantee}
\addcontentsline{toc}{subsection}{5.1 A Finite Summary with a Stated Guarantee}

Consider a handler fragment with a finite control-flow graph, known message tags, typed communication primitives and a finite sound set of possible endpoint roles at each call site. Local operations satisfy their declared effects. Each library call has an admitted finite-state communication summary that covers its concrete traces and can be inserted at that call site. These summaries represent internal communication sequences and loops through their own finite states. Branches with unknown conditions retain all their possible successors.

Insert the library summaries into the handler graph, connecting their entry and exit states to the caller\textquotesingle s control flow. Construct an automaton with one state per location in this expanded graph. Label edges by message send, receive, await and return actions, preserving their direction and endpoint role. Local computations give silent edges. If an operation may address several roles, include the corresponding alternatives. The construction retains the source\textquotesingle s branching and sequencing. It applies standard finite-state trace abstraction to the handler\textquotesingle s communication actions, including choices controlled by the handler and those supplied by its environment.

\textbf{Proposition 3 (finite communication summary).} For this fragment, construction of the automaton terminates. Every finite concrete communication trace, after values and endpoint instances are abstracted to their declared tags and roles, is a trace of the automaton.

\emph{Proof.} There are finitely many call sites, and each inserted summary has finitely many states and edges. The construction therefore visits finitely many locations in the expanded graph and adds finitely many role alternatives. At a concrete local step, its source control edge supplies a silent transition. At a primitive communication step, the effect summary supplies the action, and sound target analysis includes its actual role. Within a library call, the admitted summary supplies the corresponding path. A chosen branch is among the retained successors. Induction on the concrete execution gives the corresponding automaton path. \(\square\)

With target and library summaries supplied, the construction costs a traversal of the expanded graph plus the number of emitted alternatives. This is a useful automatic result even when data and the number of invocations are unbounded. Checking a protocol candidate then uses the recorded input/output direction, endpoint ownership and available session rules. In particular, the check must distinguish choices the handler makes from choices a client is entitled to make.

Our design draws on existing inference results with similarly explicit fragments. Spaccasassi and Koutavas infer sessions for a functional language in the ML family, under stated restrictions on communication and recursion \hyperref[ref-spaccasassi]{{[}12{]}}. Padovani reconstructs composite regular linear types, including encodings of session communication \hyperref[ref-padovani-inference]{{[}13{]}}. These results support examining a dynamically created endpoint or a handle stored in a value through its provenance and use. The inference procedure can then exploit the communication discipline that remains visible in the program.

\subsection*{5.2 Wait Dependencies}\label{wait-dependencies}
\addcontentsline{toc}{subsection}{5.2 Wait Dependencies}

A send that does not wait for a response differs from a call whose continuation requires one. For each blocking call, the compiler records all participants that may be needed to release the wait. The resulting finite relation \(W\subseteq V\times V\) is a sound over-approximation of the dependencies under examination. Here \(a\,W\,b\) means that work owned by \(a\) may wait for work owned by \(b\).

We use the standard acyclicity certificate of a natural-number rank \(r\) satisfying

\[
a\,W\,b\quad\Longrightarrow\quad r(a)>r(b).
\]

\textbf{Proposition 4 (exclusion of represented wait cycles).} If every actual blocking dependency is included in \(W\) and the rank condition holds, no execution has a circular wait consisting of those dependencies.

\emph{Proof.} A circular wait would give \(a_0Wa_1W\cdots Wa_k=a_0\). Repeated use of the rank condition yields \(r(a_0)>r(a_1)>\cdots>r(a_0)\), contradicting the strict order on natural numbers. \(\square\)

For an explicit acyclic graph, a topological traversal supplies the rank. Its construction and checking are polynomial graph operations. An encoding in quantifier-free linear integer arithmetic can also record the supplied rank as a finite arithmetic certificate. More expressive dependency families require an admitted abstraction or a separate ranking argument, with the cost and coverage of that procedure stated in its contract.

Our wait summary includes the resource dependencies relevant to its claim. A bounded queue whose sender blocks on capacity introduces a wait even if its source operation resembles a send. Locks, reply buffers and external completions can introduce others. A cycle in a conservative summary is a candidate requiring refinement or a protocol argument, since mutually exclusive paths may have contributed its edges.

\subsection*{5.3 What Makes a Request Complete}\label{what-makes-a-request-complete}
\addcontentsline{toc}{subsection}{5.3 What Makes a Request Complete}

The rank theorem excludes one cause of failure to complete. Completion also depends on terminating local work, scheduling of ready participants, delivery under the selected communication contract and discharge of external operations. These premises belong to their respective libraries and execution targets.

\textbf{Finite-batch completion lemma.} Fix a finite set of admitted invocations closed under the work spawned by that batch. Include every internal blocking dependency in a strict natural-number rank relation, covering state ownership, queue capacity and terminal cleanup as well as calls. Each invocation has a total-correctness contract for its whole continuation: if its required waits are eventually released, its handler and terminal cleanup finish. Assume fair scheduling of ready computations, eventual selection of each enabled admitted request, and eventual discharge and delivery of required external terminal outcomes. Every invocation then reaches a terminal outcome allowed by its protocol.

\emph{Proof.} At the lowest rank, an invocation has no unresolved internal prerequisite. The external-outcome and scheduling premises supply its waits and execution opportunities, so its whole-continuation contract gives a terminal outcome. Inductively, lower-ranked invocations complete and release the internal prerequisites of the next rank. The same argument then applies at that rank. The admitted set is finite, so this induction covers every invocation. The ownership argument of §4 separately gives at-most-once acceptance of each terminal outcome. \(\square\)

The whole-continuation premise covers repeated suspension and resumption. Termination of each local segment between waits would permit a handler to repeat those segments indefinitely. Request-selection fairness also concerns each queued invocation, beyond scheduling its actor. A terminal outcome can be a declared failure. Successful completion additionally requires the worker\textquotesingle s success conditions. A persistent service can continue accepting later batches after the examined batch has completed.

A native realization carries this conclusion through a further execution contract. Every fair maximal target execution must project to a fair maximal source execution with corresponding request and terminal events. Establishing that contract requires control of administrative execution, such as a well-founded measure for internal steps, and a scheduler correspondence preserving the request-selection and delivery premises. Proposition 2 supplies the finite-trace and ownership part of this argument. The maximal-execution contract supplies the completion part and must be proved for the selected runtime binding.

A compiler can therefore report an exact result about the examined structure. It may have a proved acyclic summary, a protocol theorem supporting a richer dependency pattern, or a supervised call with an explicit unresolved static obligation. A supervised call carries its timeout behavior and cleanup requirements together with the remaining static obligations.

\section*{6. When Service Order Can Change}\label{when-service-order-can-change}
\addcontentsline{toc}{section}{6. When Service Order Can Change}

Once ownership and protocol use are established, the compiler can consider additional scheduling freedom. Two requests can follow the same protocol and still compete for a value. Reordering requires a property of their computations and of what the application observes.

Suppose an actor has one available resource and two clients request it. Both orders end with no resource available, but a different client receives the successful reply. Even agreement of the final state is insufficient to make the two executions equivalent to those clients.

\subsection*{6.1 State and Labelled Replies}\label{state-and-labelled-replies}
\addcontentsline{toc}{subsection}{6.1 State and Labelled Replies}

Fix the payload of each request occurrence and retain its unique identity. Write its deterministic handler as

\[
h_i(s)=\bigl(r_i(s),u_i(s)\bigr),
\]

where \(r_i\) returns its reply and \(u_i\) updates the state. Suppose every relevant intermediate state admits the requests under consideration. For distinct occurrences \(i,j\), require

\[
\begin{aligned}
u_j(u_i(s))&=u_i(u_j(s)),\\
r_i(s)&=r_i(u_j(s)),\\
r_j(s)&=r_j(u_i(s)).
\end{aligned}
\]

The first equation concerns state. The other two preserve the answer associated with each request. We specialize the established state-and-return-value notion of commutativity described by Koskinen and Bansal to these labelled actor requests \hyperref[ref-commutativity]{{[}14{]}}.

\textbf{Proposition 5 (permutation-invariant outcomes).} If these equations hold for every pair in a finite batch and every reachable intermediate state, all service permutations yield the same final state and the same request-indexed reply map.

\emph{Proof.} Exchanging two adjacent requests preserves the state after the pair by the first equation. The remaining equations preserve both labelled replies. The unchanged suffix therefore begins in the same state. Every permutation is a product of adjacent exchanges, so repeated application gives the result. \(\square\)

The observation in this proposition consists of state and the labelled reply map. Reply chronology and elapsed time remain properties of the selected schedule. For a handler that increments a counter and returns the old value, the updates commute but the replies do not satisfy the displayed equations. The compiler can consequently retain the valid state law without using it to justify a stronger client-visible equivalence.

For a positive example, independent workers can fill distinct slots of a result map and receive a fixed acknowledgement. Updates to different slots commute, and their acknowledgements are unchanged by the other update. The lemma then permits any arrival order for those completions. A later computation reads the result map under its own combination rule.

A probabilistic handler adds a sampling obligation to this scheduling question. Reordering calls to a shared random-number generator can assign different stream positions to the labelled requests. Proposition 5 covers the deterministic state and reply observations in its premises. Preserving a stochastic joint law and its parameter derivatives requires a corresponding probabilistic argument. An invocation-indexed sampling construction or a proved coupling can justify a particular reordering. Section 9.3 uses a fixed sampling plan so that result arrival order can change without silently changing the estimator.

\subsection*{6.2 Dependency Width and Placement}\label{dependency-width-and-placement}
\addcontentsline{toc}{subsection}{6.2 Dependency Width and Placement}

The relevant parallelism is the independence that remains after data, control and resource dependencies are included. For a finite acyclic dependency graph, reachability defines a partial order. Its \emph{dependency width} is the largest number of pairwise incomparable events. Dilworth\textquotesingle s theorem identifies this number with the minimum size of a chain cover \hyperref[ref-dilworth]{{[}15{]}}.

This gives a useful structural measure for compilation. A chain describes ordered work, while an antichain identifies work whose order is unconstrained by the retained relation. Placement must still account for readiness, target capacity and the resources used by those events. A collection of pending requests is therefore distinct from a proof that their handlers can execute simultaneously.

The coordinator in §8 can construct both worker requests before either result is available. Their data and resource contracts justify independent execution, while the combination step waits for both replies. Its continuation marks the return from parallel work to dependent control. If that result selects a further batch, the compiler analyzes the next dependency boundary in the same way. Each region can then be placed or specialized while preserving the order on which the surrounding computation depends.

\section*{7. Proofs at the Communication Boundary}\label{proofs-at-the-communication-boundary}
\addcontentsline{toc}{section}{7. Proofs at the Communication Boundary}

A distributed realization must connect the logical conversation to actual memory and transport operations. A request can transfer a buffer, a right to reply or evidence about a computed value. The receiver needs a justified account of what becomes available when it accepts that request.

Our \emph{Decidable by Construction} paper introduces Iris, Actris and Aneris as sources of reusable proofs \hyperref[ref-dbc]{{[}6{]}}. Their role here can be made more specific. We use Iris\textquotesingle s ownership and invariant reasoning \hyperref[ref-iris]{{[}16{]}}, Actris\textquotesingle s dependent separation protocols \hyperref[ref-actris]{{[}17{]}}, and the distributed interpretation supplied by Aneris \hyperref[ref-aneris]{{[}18{]}}. A compiler binding supplies the correspondence between the operations those rules describe and the Clef operations being realized.

\subsection*{7.1 A Message Can Transfer a Proof Resource}\label{a-message-can-transfer-a-proof-resource}
\addcontentsline{toc}{subsection}{7.1 A Message Can Transfer a Proof Resource}

The send and receive rules in Actris 2.0, §4.2, Figure 5, advance an endpoint protocol while transferring the assertion associated with a message \hyperref[ref-actris]{{[}17{]}}. The assertion can describe values, ownership or further communication capabilities. The paper\textquotesingle s examples include transferring a heap cell and later receiving it back under a changed value assertion.

In schematic notation for our binding, let \(\mathsf{ep}(c,\Pi)\) mean ownership of endpoint \(c\) at protocol \(\Pi\), and let \(P(v)\) describe the resource accompanying value \(v\). Separating conjunction \(*\) combines resources whose ownership can be composed. The relevant rule shapes are

\[
\begin{aligned}
&\{\mathsf{ep}(c,\mathsf{Send}(P,\Pi))*P(v)\}\\
&\qquad\mathsf{send}(c,v)\\
&\{\mathsf{ep}(c,\Pi(v))\},
\end{aligned}
\]

and

\[
\begin{aligned}
&\{\mathsf{ep}(d,\mathsf{Recv}(P,\Pi_r))\}\\
&\qquad v\leftarrow\mathsf{recv}(d)\\
&\{P(v)*\mathsf{ep}(d,\Pi_r(v))\}.
\end{aligned}
\]

The sender supplies the message resource, and the receiver obtains it through the corresponding protocol step. Each side continues with the protocol determined by the received or sent value. These are compiler-facing schemata for the cited rules, with the concrete channel interpretation supplied by the binding.

For an actor request, \(P(v)\) can include the payload invariant and the server\textquotesingle s authority to produce the reply for invocation \(q\). The caller retains the dual endpoint and the right to resume its own continuation. Those are distinct resources. The handler\textquotesingle s private state remains with that handler while communication transfers the resources named by the message.

An exclusive reply authority supports at-most-once use. Actris\textquotesingle s affine reasoning also permits resources to be unused, so an eventual reply needs an additional completion argument. This is precisely the separation used in §§4--5. The communication proof establishes resource-safe protocol steps, and the progress contract supplies the conditions under which the required steps occur.

The handler itself supplies the middle step of this derivation. Write \(\mathsf{State}(s)\) for its exclusive state assertion, \(\mathsf{Req}(q,p)\) for the resources accompanying a request, and \(\Psi(q,p,v)\) for the assertion delivered with its reply. An admitted handler has a specification of the form

\[
\begin{aligned}
&\{\mathsf{State}(s)*\mathsf{Req}(q,p)\}\\
&\qquad v\leftarrow\mathsf{handle}(p)\\
&\{\exists s'.\
  \mathsf{State}(s')*\Psi(q,p,v)
  *\ulcorner\mathsf{Step}(s,p,v,s')\urcorner\}.
\end{aligned}
\]

Here the brackets \(\ulcorner\cdot\urcorner\) embed the ordinary state-transition proposition as a pure assertion. Receiving the request supplies \(\mathsf{Req}\) and its reply endpoint. Framing retains that endpoint while the handler establishes its postcondition. Sending the reply transfers \(\Psi\) while framing the successor state. If the handler suspends, its frame predicate retains these same assertions until the resume rule restores them. The proof therefore follows the actual receive, compute, suspend if needed, and reply sequence.

The endpoint in these rules is a binary session endpoint. A mailbox with several producers uses an additional queue or delegation invariant to combine their access. Its construction must preserve the independent client interfaces represented by the coexponential pool.

For a mechanized binding, the operational semantics assigns a meaning to Clef\textquotesingle s dispatch, send, receive, capture and resume operations. Proofs of those operations must establish the endpoint and frame rules used here. The handler triples above express partial correctness. The total-correctness premise in §5.3 is an additional contract on the whole handler. Rocq can check these arguments once the operational interpretation and proofs are supplied, while the compiler retains their instantiations at the corresponding program operations.

\subsection*{7.2 From Reliable Channels to an Unreliable Network}\label{from-reliable-channels-to-an-unreliable-network}
\addcontentsline{toc}{subsection}{7.2 From Reliable Channels to an Unreliable Network}

Our distributed binding needs to account for packets that can be lost, duplicated or reordered. The reliable-communication work of Gondelman and colleagues supplies a useful construction in Aneris \hyperref[ref-rclib]{{[}19{]}}. It verifies a channel library over the User Datagram Protocol (UDP) and builds remote procedure call (RPC) and replicated storage components against the library\textquotesingle s specifications.

Its session escrow argument is particularly relevant. A transmitted message has a duplicable indexed witness, while the receiving endpoint carries exclusive protocol state. Advancing that state releases the associated resource once. The proof can therefore admit duplicated packets without duplicating the logical ownership transferred by an accepted message. Figure 6 of that paper makes this distinction explicit.

For our binding, the protocol index connects a received packet to the next admissible interaction. The application invocation identity connects that interaction to a pending caller. Both survive encoding and decoding, allowing the receiver to match each result to its intended invocation even when several results are outstanding.

The RPC specification in §5 of the same work retains the request\textquotesingle s logical data in the returned postcondition \hyperref[ref-rclib]{{[}19{]}}. Its proof composes the channel send and receive rules with the application\textquotesingle s handler specification. In our worker example, that logical data includes the job identity, input version and task index. The returned assertion therefore describes the result of the requested task, which is the assertion the waiting continuation needs.

The verified channel specifications establish partial-correctness properties of the supported communication operations. A target\textquotesingle s eventual-delivery assumptions, connection lifetime and failure behavior supply the premises for a completion claim. Restart, authentication and external side effects each use their corresponding contracts. Our compiler records these premises with the channel instance so that the application proof refers to the communication service actually selected.

\subsection*{7.3 BAREWire and the Received Meaning}\label{barewire-and-the-received-meaning}
\addcontentsline{toc}{subsection}{7.3 BAREWire and the Received Meaning}

Our BAREWire boundary connects message structure to native storage and transfer. For an exact value mapping, its central value condition is

\[
\mathsf{decode}(\mathsf{encode}(v))=v
\]

on the admitted source domain. The associated obligations include valid tags, covered widths, initialized ranges, endpoint identifiers and the lifetimes of referenced storage. If the mapping changes coordinates or precision, the boundary carries that stated relation and its error bound.

Logical ownership transfer can accompany copying or serialization. The sender\textquotesingle s storage and the receiver\textquotesingle s reconstructed storage must each realize the corresponding assertion. A shared-memory realization can instead transfer or share access under its lifetime and permission contract. This permits unboxed representations while retaining the distinction between a value\textquotesingle s bytes and the authority to use the resource they designate.

The compiler-facing composition is concrete. The encoder realizes the send-side value and protocol event. The communication library relates that event to an accepted receive. The decoder realizes the receive-side value and resource assertion. The continuation then consumes the corresponding completion. Each component has a local premise that refers to the same invocation and interpretation.

\subsection*{7.4 Composing Exact Representations}\label{composing-exact-representations}
\addcontentsline{toc}{subsection}{7.4 Composing Exact Representations}

A result can pass through a worker\textquotesingle s native layout, a wire encoding and the coordinator\textquotesingle s receiving layout. Checking every pair of formats independently would repeat much of the same reasoning. We use the common-carrier discipline developed in Tarau\textquotesingle s work on isomorphic encodings to compose these boundaries from established conversions \hyperref[ref-tarau]{{[}20{]}}. Our \emph{Fixed-Point Scaffolding} paper applies this discipline to native representation interfaces \hyperref[ref-fps]{{[}7{]}}.

Let \(V\) be an admitted value domain. For each encoding \(e_i\), take its canonical image \(R_i=e_i(V)\) and a decoder \(d_i:R_i\to V\) satisfying \(d_i\circ e_i=\mathrm{id}_V\). These domains exclude invalid tags and uncovered values. Define

\[
c_{ij}=e_j\circ d_i:R_i\longrightarrow R_j.
\]

The maps \(c_{ij}\) form a groupoid on the admitted representations, with

\[
\begin{aligned}
c_{ii}&=\mathrm{id}_{R_i},\\
c_{jk}\circ c_{ij}&=c_{ik},\\
c_{ji}\circ c_{ij}&=\mathrm{id}_{R_i}.
\end{aligned}
\]

These equations follow directly from the checked round trips. Every \(x\in R_i\) has the form \(e_i(v)\), so \(e_i(d_i(x))=x\). Thus \(e_i\) and \(d_i\) are inverse on the stated domains. Substituting \(c_{ij}=e_j\circ d_i\) and cancelling \(d_j\circ e_j\) gives the composition equation. Identity and inverse equations follow by the same cancellation, and function composition supplies associativity.

The groupoid records exact changes of value representation. To realize a computation \(h:V\to V\), an operation \(h_i:R_i\to R_i\) also needs the law

\[
d_i\circ h_i\circ e_i=h
\]

on its admitted inputs. This is the connection between a codec proof and a proof about the operation that uses the codec. Distinct input and output domains use their respective encoders and decoders. A concrete native instruction must satisfy that operation law, while ownership and protocol evidence establish the authority to access or transfer the represented resource.

For the distributed request, these laws let the receiver recover the task\textquotesingle s value through several exact encodings while the protocol retains its invocation and acceptance rights. A conversion that changes precision instead supplies its stated value relation and error bound. Section 9 composes those quantitative contributions with the worker\textquotesingle s result contract. The graph retains which boundary used which law, so a format change identifies the dependent proof applications as well as the bytes to regenerate.

\section*{8. A Distributed Request from Admission to Completion}\label{a-distributed-request-from-admission-to-completion}
\addcontentsline{toc}{section}{8. A Distributed Request from Admission to Completion}

Consider a client \(C\), a coordinator \(K\) and two worker services \(W_1,W_2\). The client requests one calculation. The coordinator divides it into two named tasks, obtains their results and returns one combined answer. Each worker owns its state and serves requests through its admitted stateful protocol.

The example connects the local coexponential construction to a wider distributed program. Each service can use the worker and state-transfer laws from §3. The coordinator\textquotesingle s relationship with several services additionally uses a composition proof over its pending requests and their endpoints.

\subsection*{8.1 Identities and Outstanding Work}\label{identities-and-outstanding-work}
\addcontentsline{toc}{subsection}{8.1 Identities and Outstanding Work}

The coordinator assigns a fresh job identity \(j\) in generation \(g\) and creates two request identities \((g,j,1)\) and \((g,j,2)\). The job binds an immutable input version, so both replies refer to the selected calculation. Each request has a one-shot reply endpoint. The coordinator\textquotesingle s state records a pending set \(D=\{1,2\}\) and an initially empty result map \(R\).

Accepting a terminal reply for task \(i\) requires the matching job and generation, an admitted protocol step and \(i\in D\). Under exclusive ownership of the coordinator state, one serialized acceptance transition checks these conditions, consumes the task\textquotesingle s acceptance authority and performs

\[
(D,R)\longmapsto
\left(D\setminus\{i\},\,R[i\mapsto v_i]\right).
\]

The operation consumes that task\textquotesingle s acceptance authority. The network can deliver another copy of the same bytes, but there is no second pending entry to consume. A message from an earlier generation also fails to identify a live invocation.

For distinct task identities, these acceptance updates commute. Each writes a different slot and removes a different pending element. When both workers succeed, \(D\) becomes empty and the coordinator evaluates a specified function \(F(R[1],R[2])\). For the same worker outcomes, the result map is independent of arrival order by Proposition 5. The combination function retains its own computation contract. A floating-point reduction, for example, can use a fixed index order even when replies arrive in either order.

If one worker returns a failure, the coordinator still accounts for the other outstanding endpoint. In this example it drains both terminal outcomes before returning a failure to the client. An alternative cancellation policy requires a law that disposes of the remaining protocol resources and any late replies.

\subsection*{8.2 Suspension and Wait Structure}\label{suspension-and-wait-structure}
\addcontentsline{toc}{subsection}{8.2 Suspension and Wait Structure}

The coordinator can issue both requests before awaiting their outcomes. Its frame retains \(g,j,D,R\), the outstanding endpoints and its authority to reply to \(C\). Workers may compute concurrently, although each individual worker serializes the uses of its own state.

\begingroup\floatplacement{figure}{htbp}\setkeys{Gin}{keepaspectratio}\captionsetup{labelformat=empty,font={small,sf},margin=1em,justification=raggedright,singlelinecheck=false}

\begin{figure}
\centering
\includegraphics[width=0.8\linewidth,height=0.4\textheight]{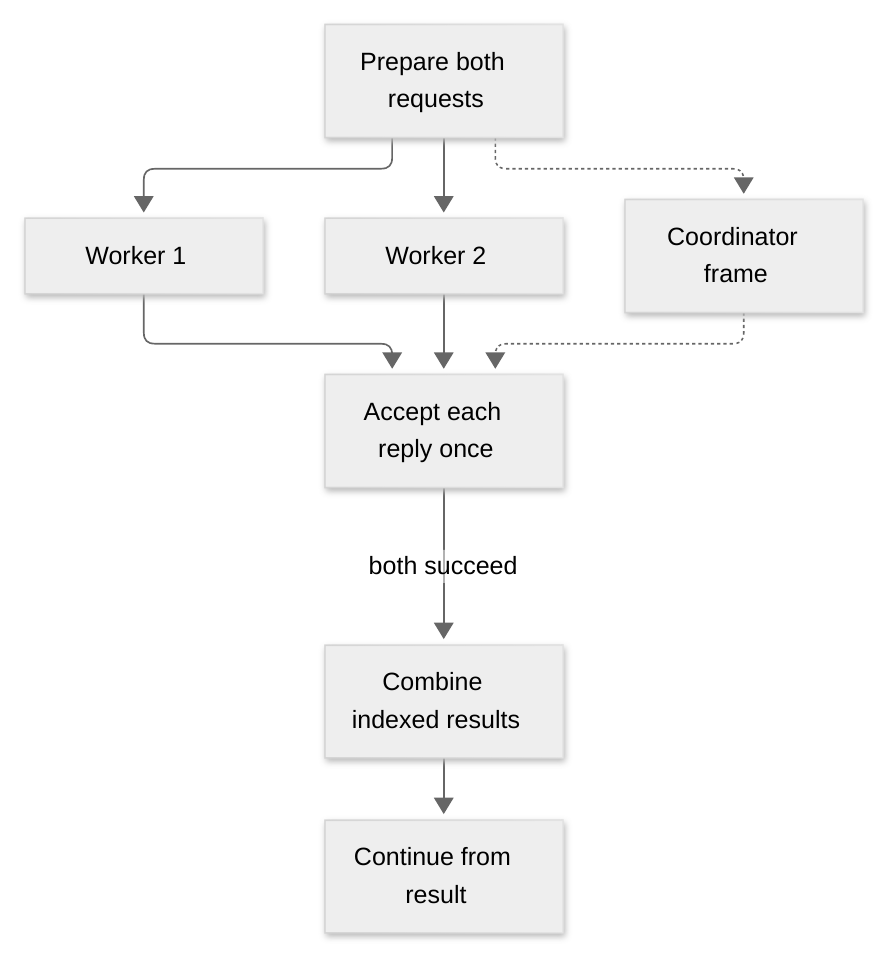}
\caption{\textbf{Figure 2.} The coordinator frame retains a suspended computation\textquotesingle s state and reply rights. Solid arrows trace requests, replies and subsequent computation; dashed arrows show retained state. Each reply is accepted at most once; successful results are combined by task identity.}
\end{figure}

\endgroup

The blocking dependencies are

\[
C\longrightarrow K,\qquad
K\longrightarrow W_1,\qquad
K\longrightarrow W_2.
\]

Ranks \(r(C)=2\), \(r(K)=1\) and \(r(W_1)=r(W_2)=0\) certify their acyclicity. The worker reply operation in this example does not require the coordinator to service another ordinary request. Any transport backpressure is discharged by the admitted channel contract or included as an additional dependency.

The rank certificate and the pending-set invariant address different failures. The first excludes a represented blocking cycle. The second prevents a result from satisfying two outstanding tasks or an old reply from resuming a new invocation. The continuation proof preserves both the state of that invariant and the code that will consume the next result.

\subsection*{8.3 Location Transparency and Direct Communication}\label{location-transparency-and-direct-communication}
\addcontentsline{toc}{subsection}{8.3 Location Transparency and Direct Communication}

A worker becomes a client when it requests another service. Each destination supplies its own \(\mathsf{Srv}(A)\) interface, with the corresponding clients represented by \(\mathsf{Pool}(A^\perp)\). The pool collects the logical participants in that service\textquotesingle s protocol. Those participants can run in one process, on separate processors or across a network. A worker can therefore address another worker directly through its typed actor reference. The supervision relationship determines lifecycle responsibility, while the endpoint relationship determines who communicates with whom.

Our location-transparent reference separates the service identity and protocol from the binding that reaches its current placement. Each admitted invocation obtains its own client endpoint and reply identity, so several holders of a service reference retain distinct conversations. A local binding can pass native values under a lifetime contract. A remote binding uses the admitted BAREWire mapping and communication protocol. In either case, the caller retains its continuation frame and awaits the result of the same identified invocation. The destination owns its state and the authority to produce that invocation\textquotesingle s reply. Direct communication can thus cross a placement boundary while preserving the endpoint and resource relationships established in §7.

For the example above, moving \(W_2\) to another node changes its delivery and execution binding. The coordinator still requests task 2 and accepts its result into slot 2. The joint graph retains the service generation, invocation identity, endpoint roles, waits and storage uses that this binding must preserve. A placement refinement relates the binding\textquotesingle s accepted requests, state transitions and replies to those of the source service. A worker\textquotesingle s reply is delivered to its caller through that binding, independently of their supervision ancestry.

Location transparency gives the application a stable protocol across admitted placements. Each placement also supplies its own admission, ordering, delivery and failure premises, as §7.2 requires. These determine whether a request can complete and which failures its caller can observe. A replacement service generation must account for outstanding requests before accepting their completion as current. The coexponential supplies the stateful interaction structure, and the placement and channel proofs establish how a particular local or remote execution realizes it.

The structure also makes resource limits visible. The example has two outstanding tasks and a known-size result map. Its frame extent, queue admission and reply capacity can be checked before execution. A service with a variable number of clients supplies an admission bound or a resource invariant that governs its growth.

\section*{9. A Solver Service with a Quantitative Contract}\label{a-solver-service-with-a-quantitative-contract}
\addcontentsline{toc}{section}{9. A Solver Service with a Quantitative Contract}

The protocol so far preserves who receives each result and how the participating resources are used. A useful service contract also states what the returned value means. A solver worker provides a concrete case in which mathematical bounds and communication proofs contribute to the same result.

Our "Adaptive Domain Models" account uses elimination-growth bounds to inform adaptive arithmetic choices \hyperref[ref-adm]{{[}21{]}}. "Decidable By Construction" and "Fixed Point Scaffolding" develop the associated verification and preservation contracts \hyperref[ref-dbc]{{[}6{]}} \hyperref[ref-fps]{{[}7{]}}. We apply that idea here to a worker solving \(Mx=b\) in declared dimensionless coordinates. The value contract records the matrix and right-hand side, the chosen algorithm, the output interpretation and the admitted error budget.

\subsection*{9.1 A Growth Bound as a Worker Premise}\label{a-growth-bound-as-a-worker-premise}
\addcontentsline{toc}{subsection}{9.1 A Growth Bound as a Worker Premise}

We use an elimination-growth bound to select candidate factor ranges for the solver service. For exact no-pivot factorization \(B=LU\), with unit lower-triangular \(L\), Urschel defines

\[
\rho(B)=\max\left\{
\|L\|_{\max},
\frac{\|U\|_{\max}}{\|B\|_{\max}}
\right\}.
\]

We apply Theorem 1.2 in Urschel\textquotesingle s paper to a fixed nonsingular real \(n\times n\) matrix \(M\) and a Haar-distributed orthogonal \(Q\). For \(0<\delta<1\), the theorem gives \hyperref[ref-urschel]{{[}22{]}}

\[
\Pr\!\left[\rho(QM)>T(n,\delta)\right]<\delta,
\qquad
T(n,\delta)=\frac{14n^{3/2}\sqrt{\log(en)}}{\delta}.
\]

The transformed system \((QM)x=Qb\) has the same exact solution. If \(H\ge\|QM\|_{\max}\) has been established, the good event gives

\[
\|L\|_{\max}\le T(n,\delta),
\qquad
\|U\|_{\max}\le T(n,\delta)H.
\]

These are factor-range premises for the worker implementation. Its realization also supplies the arithmetic analysis connecting exact and computed factors, bounds for the other intermediate values, and the effect of forming the transformed inputs. A sampler contract connects the finite implementation to the distribution required by the mathematical theorem.

The resulting output can be related to the solution through a residual argument. For example, if \(K\ge\|M^{-1}\|\) and \(\zeta\ge\|b-M\widehat x\|\) are established in compatible norms, then

\[
\|\widehat x-x\|\le K\zeta.
\]

The residual bound includes the error of its actual computation. This connects a returned value to its specified equation. Deterministic storage safety uses covered capacities or guards before operations that could exceed them, together with an admitted failure or fallback path.

The growth result consequently has a specific engineering use. It informs the candidate factor ranges and arithmetic configuration. The realized algorithm and residual contract establish the returned accuracy. The worker\textquotesingle s protocol carries the resulting success or failure as an ordinary typed outcome.

\subsection*{9.2 Accuracy at the Completed Join}\label{accuracy-at-the-completed-join}
\addcontentsline{toc}{subsection}{9.2 Accuracy at the Completed Join}

Interpret the task and completion events in one admitted probabilistic execution model. Its sampling and scheduler assumptions must justify each worker bound for the corresponding invocation.

Fix a finite admitted task set. Let \(\mathsf{Done}\) mean successful completion of the reduction after exactly one successful reply has been accepted for every task in that set. Let worker \(i\) have reference contribution \(z_i\). On success it produces \(\widehat z_i\), under a contract

\[
\Pr\!\left[
\mathsf{Success}_i
\ \land\
|\widehat z_i-z_i|>\varepsilon_i
\right]\le\delta_i.
\]

Suppose the boundary supplies a received value \(\widetilde z_i\) with \(|\widetilde z_i-\widehat z_i|\le\eta_i\). On successful completion the coordinator\textquotesingle s specified reduction returns \(\widehat z\) with error at most \(\eta_{\mathrm{red}}\) relative to \(\sum_i\widetilde z_i\). All contributions have a common unit, and the nonnegative bounds \(\varepsilon_i\), \(\eta_i\) and \(\eta_{\mathrm{red}}\) use that unit.

\textbf{Proposition 6 (accuracy of a completed distributed result).} For a finite admitted task set, exact task identification and single acceptance at the join give

\[
\Pr\!\left[
\mathsf{Done}\ \land\
\left|\widehat z-\sum_i z_i\right|
>
\eta_{\mathrm{red}}+\sum_i(\varepsilon_i+\eta_i)
\right]
\le\sum_i\delta_i.
\]

\emph{Proof.} If the join completes and every successful worker output satisfies its error bound, repeated triangle inequalities bound the reduction error by the displayed sum. A completed result exceeding that sum therefore requires at least one worker\textquotesingle s bad-success event. The union bound gives the probability inequality. Single acceptance and task identification ensure that the sum contains exactly the specified contributions. \(\square\)

This adapts Proposition 7.2 of our \emph{Decidable by Construction} paper \hyperref[ref-dbc]{{[}6{]}} to the invocation and protocol invariants above and a reduction with an explicit error bound. Independence between worker errors is unnecessary for the union bound. If work is selected adaptively, the worker contract must hold for the selected inputs under the history on which the selection depends.

The event \(\mathsf{Done}\) is part of the statement. The bound concerns completed inaccurate outputs, while §5 supplies the premises for completion. A retry policy also records its attempts and their sampling assumptions. Reusing a sample after observing its outcome requires the corresponding conditional argument.

The coexponential protocol describes how a worker remains available across these requests. The continuation and network proofs connect its particular result to the waiting computation. The quantitative proof then makes that result useful to an application with an explicit accuracy requirement.

\subsection*{9.3 A Distributed Metric Update}\label{a-distributed-metric-update}
\addcontentsline{toc}{subsection}{9.3 A Distributed Metric Update}

Our ADM construction gives this service a further numerical task: assemble and solve a restricted natural-gradient update \hyperref[ref-adm]{{[}21{]}}. The loss derivatives describe which parameter changes improve the model; a metric supplies the comparison of their sizes used to choose an update. Karwowski, Nielsen and Staton supply a compositional score-\/-velocity metric for an admitted probabilistic program presentation \hyperref[ref-karwowski-2026]{{[}23{]}}. ADM applies that metric to selected parameter directions and derives the update and its error enclosure. Here we connect that calculation to the invocation, sampling and join contracts already established.

Fix one parameter point \(\theta\), a loss \(\ell\) and linearly independent parameter directions \(U=[u_1\ \cdots\ u_k]\). With the admitted program metric \(G_\theta\), the requested system is

\[
K_{ij}=G_\theta(u_i,u_j),\qquad b_i=D\ell_\theta[u_i],\qquad Ka=b.
\]

The result supplies the parameter direction \(Ua\); an admitted descent step uses its negative with a separate step-size and constraint-preservation rule. The job binds the model and loss versions, \(\theta\), \(U\), trace presentation, observed execution information, output metric and scales, sampling plan and arithmetic contract. These identify the mathematical calculation whose distributed realization is being admitted. We apply the geometry to a finite batch of fragments with established semantics and regularity. The persistent service protocol does not itself establish geometric semantics for arbitrary loops or recursion.

The partition of work must preserve the quantities being estimated. For the full-trace score-\/-velocity metric used here, let \(s_{r,i}\) be the derivative of execution \(r\)\textquotesingle s log trace density along \(u_i\), called its directional score. Let \(H_{r,ij}\) pair the corresponding output derivatives under the selected output metric, with positive scale \(\kappa\) as in ADM. A coarser observation requires its corresponding projected quantities and a justified estimator. A sample calculation for \(N\) executions of the full-trace presentation is

\[
\overline K_{ij}=\frac1N\sum_{r=1}^{N}
\left(s_{r,i}s_{r,j}+\kappa H_{r,ij}\right).
\]

The two directions in each term refer to the same admitted execution. A straightforward distribution partitions execution indices into disjoint worker batches. Each worker returns partial matrix sums; the coordinator accepts each assigned contribution once and divides their total by \(N\). The sampling law supplies the statistical justification, including independence assumptions wherever the selected confidence bound requires them. Direction or matrix-block partitioning instead needs a construction that preserves the required joint samples. Unequal batch sizes cannot be combined by an unweighted average of their local means.

The calibration model in ADM shows why this matters. Its scores in the offset and gain directions are \(r_0+r_1\) and \(r_1\), where \(r_0,r_1\) are independent standard-normal residuals. The correct cross entry is

\[
\mathbb E[(r_0+r_1)r_1]=1.
\]

Pairing the first score with an independently resampled gain score \(r'_1\) instead gives zero. Each worker can return correctly typed values to the correct caller while their combination estimates the wrong metric. The invocation protocol therefore carries the execution or coupling identity needed by each cross term, in addition to its task identity. Derivative observations for \(b\) retain their own exact or sampling contract under the same frozen job.

The returned matrix \(\widehat K\) and vector \(\widehat b\) also include arithmetic, reduction and boundary errors. The entrywise completed-result argument of §9.2 can be applied jointly, with its failure budgets, and converted to the norms required by ADM. For example, bounds \(|\widehat K_{ij}-K_{ij}|\le e_{ij}\) on one joint event imply

\[
\|\widehat K-K\|_2\le\|\widehat K-K\|_F
\le\left(\sum_{i,j}e_{ij}^2\right)^{1/2}.
\]

The corresponding Euclidean bound applies to vector-entry errors. The entrywise premises must include sampling error relative to \(K,b\), as well as the error of assembling their estimates. Exactly-once acceptance alone supplies none of those statistical bounds.

Once \(\lambda_{\min}(K)\ge m>0\), metric and derivative error bounds, and an enclosed solve residual have been established, the solver returns an update with ADM\textquotesingle s restricted-direction enclosure. ADM §2.4.4 makes the conditioning premise checkable: for symmetric \(\widehat K\), the reply can carry a certified lower bound \(\lambda_{\mathrm{lo}}\) on its smallest eigenvalue and the metric error bound \(\varepsilon_K\). A positive lower enclosure of \(\lambda_{\mathrm{lo}}-\varepsilon_K\) supplies \(m\). Singular systems require its kernel-compatibility conditions or an explicitly changed regularized objective.

The computed direction has a separate admission check. The reply can carry the derivative-error bound and certified dot-product and norm bounds required by ADM\textquotesingle s Corollary 2.2, establishing a positive descent margin for the returned \(\widehat a\). This certificate establishes first-order descent of \(-U\widehat a\) independently of metric accuracy. The conditioning certificate is used to bound error relative to the exact metric solve. A small solve residual alone establishes neither claim. The consumer binds the returned direction and each certificate to the same model, loss, parameter point, basis and job, then checks the evidence required by its declared policy. Finite-step improvement retains the separate step-size and constraint-preservation obligations.

The quantitative guarantee remains a bound on completed inaccurate results of the form in §9.2; conditioning on successful completion requires its own argument. Attempts, resampling and any adaptive selection retain their probability premises. Statistical certificates retain their joint-event premises across the estimates and bounds used in admission.

This gives the coexponential service a concrete role in adaptive computation. It maintains a reusable worker while each invocation preserves the particular observations and numerical evidence needed for one update. A suspended request retains its original model and metric version. Applying its result after the service has moved to another parameter point requires a stale-update or migration contract; otherwise the consumer rejects it or requests a new calculation. ADM owns the optimizer derivation, while the present construction establishes how its distributed contributions and completion remain associated with that derivation.

\section*{10. Keeping the Proofs Useful to the Compiler}\label{keeping-the-proofs-useful-to-the-compiler}
\addcontentsline{toc}{section}{10. Keeping the Proofs Useful to the Compiler}

The protocol is available before a particular request arrives. Its eventual service order, values and timing belong to execution. This separation follows the static/dynamic distinction developed in the ML tradition by Harper, Mitchell and Moggi \hyperref[ref-phase]{{[}24{]}}. Our compiler retains static facts long enough for them to constrain the operations selected to execute that dynamic behavior.

\subsection*{10.1 Four Tiers and Their Interfaces}\label{four-tiers-and-their-interfaces}
\addcontentsline{toc}{subsection}{10.1 Four Tiers and Their Interfaces}

Our \emph{Decidable by Construction} paper organizes verification into four tiers \hyperref[ref-dbc]{{[}6{]}}. In this application their roles are concrete.

{\def\LTcaptype{none} % do not increment counter
\begin{longtable}[]{@{}
  >{\raggedright\arraybackslash}p{(\linewidth - 4\tabcolsep) * \real{0.3333}}
  >{\raggedright\arraybackslash}p{(\linewidth - 4\tabcolsep) * \real{0.3333}}
  >{\raggedright\arraybackslash}p{(\linewidth - 4\tabcolsep) * \real{0.3333}}@{}}
\toprule\noalign{}
\begin{minipage}[b]{\linewidth}\raggedright
Tier
\end{minipage} & \begin{minipage}[b]{\linewidth}\raggedright
Evidence used for the actor construction
\end{minipage} & \begin{minipage}[b]{\linewidth}\raggedright
Compiler consequence
\end{minipage} \\
\midrule\noalign{}
\endhead
\bottomrule\noalign{}
\endlastfoot
\mbox{1.} Structural inference & Message and value types, endpoint uses, admitted session constructors and dimensional identities & Construct the typed participants and their structural relationships \\
\mbox{2.} Local generated checks & Frame extents, initialized slots, covered conversions and finite wait-rank conditions & Admit the local representation and dependency choices \\
\mbox{3.} Reusable domain and system proofs & Handler invariants, protocol resource transfer, channel refinement and solver bounds & Apply general laws to the actual requests, state and boundary operations \\
\mbox{4.} Relational proofs & Equivalence of selected execution strategies, probabilistic error composition and richer history relations & Establish the application\textquotesingle s required relationship between executions or results \\
\end{longtable}
}

Tiers 1--3 are designed to obtain their application parameters from source and graph structure. Library authors supply the general lemmas and their admitted bindings. Ordinary projects using covered operations acquire those proofs through compilation, without repeating the library\textquotesingle s protocol annotations. Tier 4 can use project annotations to select the required relation. Reusable relational constructions can progressively supply those choices from recognizable graph structure.

The tiers also share established premises. A checked buffer extent can enter a distributed resource-transfer proof. A typed request identity can be reused in a relational statement about reordered execution. A Tier 2 range result can contribute directly to the Tier 4 completed-output bound.

We use the mode and shift account in Hăvărneanu\textquotesingle s \emph{Classical Adjoint Logic} to inform these interfaces \hyperref[ref-havarneanu]{{[}25{]}}. Each translation specifies what judgment it accepts, how its premises are interpreted and what judgment it makes available in the receiving mode. Reuse preserves those dependencies, so a higher-tier proof can begin with facts already established about the same program.

For the coexponential case, the protocol theorem and its internal fixed-point reasoning remain with the logic in which they were established. The compiler imports an applicable judgment through its stated interpretation. The syntax of adjoint shifts alone does not establish cut elimination for a new combined fixed-point calculus. The design here uses explicit proof interfaces whose composition obligations can be stated independently of that larger metatheory.

\subsection*{10.2 Composing Components and Their Refinements}\label{composing-components-and-their-refinements}
\addcontentsline{toc}{subsection}{10.2 Composing Components and Their Refinements}

An engineer composes a worker with a channel and a receiving continuation. During compilation, each of those components can also acquire a different realization. We need to compose the computations along their interfaces and compose the evidence for their replacement. These are the two levels of the following construction. It uses the standard locally preordered form of a two-category, illustrated by relation inclusion in Lack\textquotesingle s account \hyperref[ref-lack]{{[}26{]}}.

Fix an environment \(\Gamma\) of assumptions and an admitted interpretation of source and target components. An interface specifies values, resources and permitted observations. A component exposes its invocation and protocol relationships through that interface. Write \(K[-]\) for a well-typed surrounding program with one component position. The admitted contexts depend on the exposed contract and are closed under typed plugging, so an accepted replacement remains eligible in the same surrounding program.

Our objects are these interfaces. A 1-cell \(f:A\to B\) is a finite typed path of admitted component operations from \(A\) to \(B\). Composition concatenates paths, and the empty path is the identity. Concurrent behavior can occur within a component and within its surrounding program. Its interpretation supplies a set \(\operatorname{Obs}_\Gamma(K[f])\) of finite visible traces, including the request identities, replies, effects and state observations declared by the contract.

For components with the same boundary, define a 2-cell \(\alpha:f\Rightarrow f'\) when

\[
\operatorname{Obs}_\Gamma(K[f'])
\subseteq
\operatorname{Obs}_\Gamma(K[f])
\qquad\text{for every admitted }K[-].
\]

Thus \(f'\) refines the permitted behavior of \(f\) under the fixed assumptions. Equality of the observation sets gives comparisons in both directions. For this categorical account, one cell records that a comparison holds. The compiler retains the particular proof, premises and artifact identities used to establish it as evidence for that cell.

\textbf{Proposition 7 (composition of contextual refinements).} These interfaces, paths and comparisons form a strict locally preordered two-category. Successive refinements compose, and refinements of interface-compatible components compose horizontally. In particular, if \(\alpha:f\Rightarrow f'\) and \(\beta:g\Rightarrow g'\), then

\[
\beta*\alpha:g\circ f\Rightarrow g'\circ f'.
\]

\emph{Proof.} Reflexivity and transitivity of inclusion supply identity comparisons and vertical composition. Closure of the admitted contexts under plugging gives

\[
\begin{aligned}
\operatorname{Obs}_\Gamma(K[g'\circ f'])
&\subseteq\operatorname{Obs}_\Gamma(K[g\circ f'])\\
&\subseteq\operatorname{Obs}_\Gamma(K[g\circ f]),
\end{aligned}
\]

by applying \(\beta\) in \(K[-\circ f']\) and \(\alpha\) in \(K[g\circ-]\). This supplies horizontal composition. Path concatenation is associative with the empty path as its unit. Each pair of parallel paths has at most one comparison cell, so horizontal composition preserves identities and vertical composition. In particular, the two orders of composing a rectangular family of comparisons give the same cell, which is the interchange law. \(\square\)

The proposition applies the standard categorical construction to our declared component interfaces. Its use in the compiler is concrete. A channel implementation and a worker realization can each establish a refinement against their own contract, then contribute those results to their composition. A particular library proof must establish the contextual comparison, using its state, resource and protocol invariants. The compiler instantiates that proved rule at the actual component boundaries.

This comparison concerns finite observable behavior. A required completion or probability guarantee also retains its admitted judgment and premises from §§5 and 9. A scheduler replacement, for example, must preserve the fairness assumptions used by the completion proof. The compiler associates these additional obligations with the same component, so a refinement of its visible behavior can be used together with the guarantees the application requires.

The metric service makes this distinction numerical. Equality of possible reply traces does not establish equality of their probabilities or the derivatives used by \(K\). Even preserving the output law can change the declared trace geometry, as when a transformation removes a parameter-dependent draw whose value is discarded \hyperref[ref-karwowski-2026]{{[}23{]}}. A replacement used by §9.3 must preserve the selected geometry through an admitted relation or re-establish the affected update contract. A target may calculate a retained source metric through a justified relation even when its own execution geometry differs.

The distinction from \(\operatorname{Para}(\mathcal C)\) is useful. Section 3.3 compares parameterized computations through reparameterization equations. Here the comparison concerns contextual behavior across admitted operational realizations. A specialized native handler can participate in both accounts after its encodings establish the common value interpretation. The graph retains the specific law used by each transformation.

\subsection*{10.3 Evidence Across Compilation and Reasoning Modes}\label{evidence-across-compilation-and-reasoning-modes}
\addcontentsline{toc}{subsection}{10.3 Evidence Across Compilation and Reasoning Modes}

A proved component comparison must remain connected to the operations that realize it. Consider a bound on a result buffer. The source proof refers to a particular allocation and invocation. Its native use refers to an address calculation and a resumed frame. The compiler must preserve that association when it lowers the operation and when it imports the bound into a protocol or relational proof.

We specialize the proof-transport account of \emph{Fixed-Point Scaffolding} §5.2 to this interaction \hyperref[ref-fps]{{[}7{]}}. Let \(C_m(p)\) describe evidence about artifact \(p\) in reasoning mode \(m\). Let \(V_p:C_m(p)\to C_n(p)\) be an admitted mode transport, and let \(T_L^m:C_m(p)\to C_m(Lp)\) transport evidence through a lowering \(L\). On the evidence and side conditions for which both routes are admitted, require

\[
T_L^n\circ V_p\ \simeq\ V_{Lp}\circ T_L^m.
\]

The equivalence \(\simeq\) states that the routes justify the corresponding target property under matched assumptions. Hăvărneanu\textquotesingle s shift adjunctions supply the proof-context discipline for applicable mode interfaces \hyperref[ref-havarneanu]{{[}25{]}}. Our compiler\textquotesingle s correspondence additionally binds those interfaces to the lowered operation. A reusable arithmetic fact can therefore support a distributed proof without losing the buffer, invocation or ownership premises that make it applicable.

\begingroup\floatplacement{figure}{htbp}\setkeys{Gin}{keepaspectratio}\captionsetup{labelformat=empty,font={small,sf},margin=1em,justification=raggedright,singlelinecheck=false}

\begin{figure}
\centering
\includegraphics[width=0.8\linewidth,height=0.4\textheight]{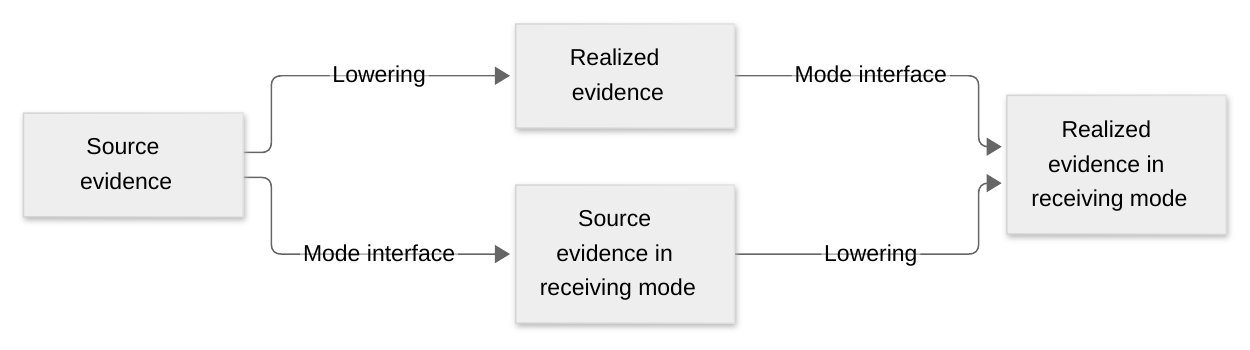}
\caption{\textbf{Figure 3.} Lowering carries evidence to a target implementation; a mode interface carries it between reasoning contexts. The two admitted paths justify the same target property under matched assumptions, although their proof terms may differ.}
\end{figure}

\endgroup

Our compilation-sheaf account organizes compatibility along the resulting stage diagrams. For a fixed admitted finite stage order, let \(E_s\) be the space of fact interpretations at stage \(s\), and let \(D_{s,t}:E_s\to E_t\) obey identity and composition laws. Interpretations identify evidence according to the declared equivalence. A compatible assignment has

\[
D_{s,t}(e_s)=e_t
\]

along the order. Once the map laws hold, checking this assignment on the immediate cover edges suffices for longer chains. For two successive edges,

\[
D_{s,u}(e_s)
=D_{t,u}(D_{s,t}(e_s))
=D_{t,u}(e_t)
=e_u.
\]

This is the finite-poset compatibility principle described in Definition 2.17 and Remark 2.19 of the sheaf account by Ayzenberg and colleagues \hyperref[ref-sheaf]{{[}27{]}}. We use the compilation stages as the indexing structure. The actor\textquotesingle s communication and waiting relationships remain within the facts and obligations being transported.

Each stage also needs evidence that its facts describe its actual code. Using the preservation contract from \emph{Fixed-Point Scaffolding} §4.2 \hyperref[ref-fps]{{[}7{]}}, write

\[
\begin{aligned}
&\operatorname{Valid}_s(p,e)\land H_{s,t}(p,e)\\
&\qquad\Longrightarrow
\operatorname{Valid}_t(L_{s,t}(p),D_{s,t}(e)).
\end{aligned}
\]

Here \(H_{s,t}\) contains the lowering\textquotesingle s particular side conditions. The compatible assignment records agreement between stage descriptions. The preservation contract establishes that the generated operation satisfies the description. Together they let a library proof remain useful from the source interaction to its native execution.

Joint constraints keep this evidence consistent across analyses. Two proof applications that use the same invocation must retain that identity, just as a layout and lifetime proof must agree on the allocation they describe. Aggregate constraints also retain every participant. Three simultaneously live buffers of four units each exceed a capacity of ten even though every pair fits. The hypergraph records the full capacity predicate and the liveness facts that justify it. A changed schedule can permit storage reuse when its proof establishes the new overlap relationship.

\subsection*{10.4 From a Frame to Native Operations}\label{from-a-frame-to-native-operations}
\addcontentsline{toc}{subsection}{10.4 From a Frame to Native Operations}

The representation, refinement and proof-transport laws have distinct roles, and each leads to a concrete compiler consequence. An exact codec establishes the value interpretation of a chosen layout. A contextual refinement admits a component replacement within its receiving interfaces. A coherent proof transport makes an established fact available at the operation that consumes it. These are the conditions our compiler uses when selecting a native realization.

Within Clef Compiler Services, Baker constructs the semantic relationships and obligations. Alex witnesses their settled consequences in portable MLIR. Composer\textquotesingle s backends realize those operations for the target, following the preservation account developed in \emph{Fixed-Point Scaffolding} \hyperref[ref-fps]{{[}7{]}}.

Our continuation realization expresses the return point, live values and dispatch state through ordinary control-flow and storage operations. The thin middle end preserves their association while admitting the final representation. A callable\textquotesingle s code value and captured environment must still refer to the same computation, and a frame\textquotesingle s live slots must remain valid at the selected resumption point.

Our graph likewise retains the message\textquotesingle s source interpretation when selecting an unboxed record or buffer. Statically resolved type bookkeeping can be erased after its consequences have been used to justify layout and execution. Dynamic state, pending identities and resources remain for as long as the conversation needs them.

This applies across native targets for central processing units (CPUs), graphics processing units (GPUs), neural processing units (NPUs) and field-programmable gate arrays (FPGAs), through their respective realization contracts. A service can coordinate accelerator work while its continuation retains the device completion and buffer lifetimes. The arithmetic fragment and its controller then have distinct target obligations connected by the same program graph.

Foreign interfaces provide a related case. A C or C++ callback that completes a request must be associated with the correct captured environment, invocation and lifetime. The binding preserves the argument and result conversions and the completion policy. Our companion paper\textquotesingle s principled marshaling account gives this boundary a concrete compiler representation \hyperref[ref-fps]{{[}7{]}}.

\subsection*{10.5 Stateful Services Across Processor Classes}\label{stateful-services-across-processor-classes}
\addcontentsline{toc}{subsection}{10.5 Stateful Services Across Processor Classes}

Our target interpretation preserves a service\textquotesingle s state transitions and labelled replies while selecting how each worker executes. Ownership governs access and release. Placement determines whether values occupy registers, cacheable memory, device buffers or storage local to an accelerator. An actor\textquotesingle s bounded state can therefore benefit from a compact layout and nearby computation. A suspension preserves its live values in the frame representation admitted for that target.

Processor-cache coherence and cached-result validity have separate roles. The memory contract establishes which writes a consumer can observe. The incremental contract establishes whether a previous result still follows from its inputs and effects. Our graph retains both dependencies, allowing locality improvements and result reuse to be justified against their respective premises.

{\def\LTcaptype{none} % do not increment counter
\begin{longtable}[]{@{}
  >{\raggedright\arraybackslash}p{(\linewidth - 4\tabcolsep) * \real{0.3333}}
  >{\raggedright\arraybackslash}p{(\linewidth - 4\tabcolsep) * \real{0.3333}}
  >{\raggedright\arraybackslash}p{(\linewidth - 4\tabcolsep) * \real{0.3333}}@{}}
\toprule\noalign{}
\begin{minipage}[b]{\linewidth}\raggedright
Processor class
\end{minipage} & \begin{minipage}[b]{\linewidth}\raggedright
Realization and engineering value
\end{minipage} & \begin{minipage}[b]{\linewidth}\raggedright
Preservation obligations
\end{minipage} \\
\midrule\noalign{}
\endhead
\bottomrule\noalign{}
\endlastfoot
Hosted CPU, including SIMD & Cache-aware state layout and vector execution of independent work within a handler & Lane effects, aliases, arithmetic and memory ordering \\
GPU with SIMT execution & Cooperating threads execute a request\textquotesingle s data-parallel region & Thread synchronization, buffer lifetime and device completion \\
Tiled NPU & Local operands and streams connect compatible compute regions & Tile admission, transfer order, capacity and result identity \\
FPGA & Request arbitration, owned state and a worker datapath form a spatial service & State and reply correspondence, backpressure and reset behavior \\
Single-core MCU & Bounded continuation frames connect device events to waiting services & Register effects, interrupt delivery, frame bounds and supervision capacity \\
\end{longtable}
}

Single instruction, multiple data (SIMD) execution applies a vector operation to several elements. Single instruction, multiple threads (SIMT) execution supplies a group of threads with their own state and control. In both cases, our compiler obtains parallel work from the handler\textquotesingle s admitted dependencies and effects. For an array update, the handler proof can assign disjoint output elements to separate lanes or threads and reassemble ownership before publishing the successor state. This permits acceleration within one state transition while preserving the order between requests. For a GPU, the realization also supplies the synchronization and asynchronous-completion conditions of the target \hyperref[ref-gpu-execution]{{[}28{]}}. A reduction retains its stated arithmetic order or uses a separately established equivalence or error bound.

Our tiled NPU realization binds a request to compatible compute regions and to the transfers connecting them. Architectures such as AMD\textquotesingle s AI Engine expose tile-local memory, direct memory access (DMA) and stream connections \hyperref[ref-aie]{{[}29{]}}. These provide concrete places for the compiler to retain operands across successive operations and overlap admitted transfer with computation. The controller carries the service state and request identity while the tiles perform the selected work. Queue capacity and transfer waits enter the same resource and dependency account as the service\textquotesingle s other waits.

Static dependency regions admit a fixed tile and route configuration. A result that selects subsequent work returns to the controller for the next admitted region. Incremental updates retain unchanged inputs that a join still needs alongside its recomputed inputs. The service can therefore reuse both configured computation and valid results while preserving the request\textquotesingle s complete input relationship.

An FPGA permits the service control and its computation to be expressed together as a spatial circuit. A bounded client-pool interface supplies ingress queues and arbitration. Registers or local memories hold the owned state. A worker datapath transforms that state and produces the selected client\textquotesingle s reply, while pending-request state records where an interaction will continue. This gives the initialization, service and finalization construction a direct hardware interpretation. Its refinement obligation relates accepted requests, state commits and replies to the source transitions, allowing internal clocked steps between those events. Pipelining several requests requires an invariant preserving those same state and reply relationships. Admission, backpressure and reset complete the contract for a finite device.

For a single-core microcontroller unit (MCU), concurrent services can share one processor while awaiting device events. Memory-mapped input/output (MMIO) accesses are effectful protocol operations with specified widths, ordering and register behavior. The compiler preserves required reads, writes and acknowledgements when specializing a handler. Interrupt bindings make the intended continuation ready, with the frame and resource bounds established before execution. This supports a complete device-oriented unikernel through the same service, continuation and ownership account used for hosted distributed applications.

Islaris provides a machine-code verification route under supported instruction-set semantics, including MMIO interactions governed by an explicit device protocol \hyperref[ref-islaris]{{[}30{]}}. Applying that route to this service requires a correspondence connecting verified device events to the pending invocation, state ownership and continuation resumption, as described by the target obligations in DBC §6.3 \hyperref[ref-dbc]{{[}6{]}}. A different MCU requires the applicable instruction and peripheral semantics.

\subsection*{10.6 Execution, Supervision and Scheduling}\label{execution-supervision-and-scheduling}
\addcontentsline{toc}{subsection}{10.6 Execution, Supervision and Scheduling}

Our runtime separates dispatch, supervision and worker execution. Ariel supplies scheduling for both Prospero system actors and Olivier worker actors. Prospero governs placement, actor lifecycles and recovery policy. Olivier supplies worker execution with its state and resource ownership. Prospero\textquotesingle s supervision guarantees depend on Ariel\textquotesingle s dispatch and reserved control-plane capacity, so a supervisor remains executable when ordinary worker traffic saturates its admitted resources.

For a worker, Ariel resumes a ready continuation. A turn runs from that resumption to suspension, completion or fault. The fragment in §4 retains actor-state authority across suspension, so another request cannot acquire it while that handler awaits a reply. When a turn exhausts its admitted budget, Prospero can retire the actor, restart it with a fresh arena or escalate the fault. These remedies operate at actor-lifecycle granularity. Ariel supplies the execution capacity through which the selected policy takes effect.

\begingroup\floatplacement{figure}{htbp}\setkeys{Gin}{keepaspectratio}\captionsetup{labelformat=empty,font={small,sf},margin=1em,justification=raggedright,singlelinecheck=false}

\begin{figure}
\centering
\includegraphics[width=0.8\linewidth,height=0.4\textheight]{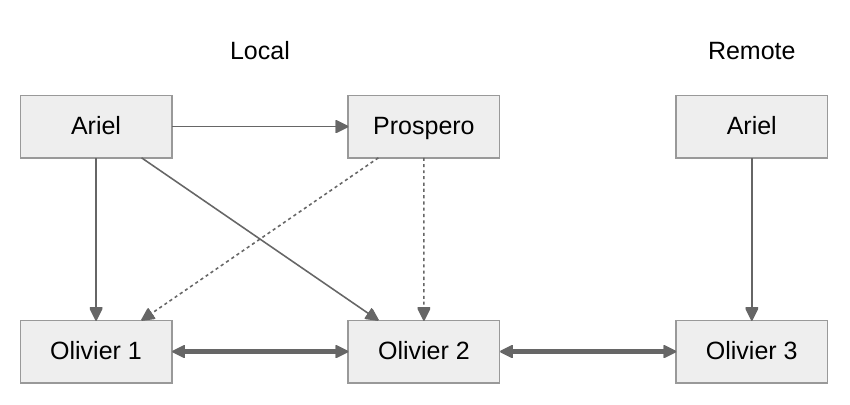}
\caption{\textbf{Figure 4.} Ariel schedules actors; Prospero governs placement, lifecycles and recovery; Olivier workers execute with owned state and resources. Thin arrows denote dispatch, dashed arrows supervision, and bold links typed worker messages. Local and remote domains have separate schedulers.}
\end{figure}

\endgroup

The message edges in Figure 4 connect workers directly. Delivery records a message at its typed destination. The actor\textquotesingle s protocol and current state determine when that message enables a turn. An accepted reply for a live pending invocation makes its corresponding continuation ready for Ariel. Prospero\textquotesingle s lifecycle decisions apply to those actors independently of the path their application messages take. Each node retains its own dispatch authority when Prospero coordinates services across machines. On a hosted target, the operating system may pre-empt carrier threads while Olivier and Ariel preserve the actor\textquotesingle s logical turn discipline. Dispatch and delivery assumptions remain premises of the progress argument.

On our single-core unikernel profile, the retained continuation frame also serves as the task, and its interrupt-to-continuation binding supplies dispatch. Cooperative scheduling therefore reuses the suspension and resumption already required by the program. Timers, watchdogs and supervision use reserved control-plane capacity and exception priorities, armed before the first turn. The same continuation representation preserves a computation across a wait and supplies the native point at which Ariel returns it to execution.

Retirement also accounts for work already using the actor\textquotesingle s resources. Prospero\textquotesingle s sentinel records the validity of an actor reference. Releasing backing storage additionally requires every outstanding device or foreign use to complete, or a target-specific revocation guarantee that establishes the end of access. The need to order release after asynchronous uses is explicit in GPU memory-management contracts \hyperref[ref-gpu-execution]{{[}28{]}}. Our realization therefore separates closing admission, retiring an identity and reclaiming its storage. A fresh generation can use a new arena while earlier buffers remain reserved under their completion contracts. Admission bounds include that still-live storage. Generation checks protect reply identity, and the retained lifetime protects the memory that outstanding work can reach.

\subsection*{10.7 Differential Compilation}\label{differential-compilation}
\addcontentsline{toc}{subsection}{10.7 Differential Compilation}

The same dependency discipline supports differential compilation. Runtime incremental computation validates cached values against their inputs. Here the compiler validates proofs, layouts and lowered fragments against the revised program and their supporting premises. Changing a reply representation affects its encoder, decoder and dependent handler bindings. Changing a call target affects the protocol application and wait summary. Changing a solver\textquotesingle s precision or sampler affects its arithmetic or probability premises.

For the distributed metric update, these dependencies also include the trace presentation, observation, output scales, tangent basis, parameter point and sampling or coupling plan. A sampler, foreign callback or scheduling change can invalidate dependent geometric evidence even when the message layout is unchanged. Generic worker laws can remain reusable while the numerical matrix, measurements and certificates bound to a particular job must be recomputed or explicitly transported.

Our differential compilation account uses these dependencies to identify the affected program fragments. It can retain a proof or lowered fragment when its supporting premises, interpretation and operation correspondence remain valid. A Huet zipper provides a focused graph traversal with its surrounding context, as developed for segmented witnessing in \emph{Fixed-Point Scaffolding} \hyperref[ref-fps]{{[}7{]}}. The focus identifies the fragment under change, and the context records what must still hold when its replacement is assembled.

The two-category account gives this reuse a precise boundary. Suppose the focused component \(f\) is replaced by \(f'\) under the same assumptions and exposed contract. A proved comparison \(f\Rightarrow f'\) supports the corresponding replacement in an admitted context \(K[-]\). When an edit changes the interface or a supporting assumption, revalidation expands through the affected joint constraints before the replacement is composed. The combined result below connects this contextual comparison to realized code and to the evidence that can be retained.

\textbf{Proposition 8 (coherent realized replacement with differential reuse).} Fix the admitted environment \(\Gamma\), a context \(K\) and same-interface components \(f,f'\). Let \(\pi\) establish a safety judgment \(\Phi(K[f])\) that is preserved under inclusion of finite observable traces, and let a checked comparison establish \(f\Rightarrow f'\). Suppose the revised source satisfies the additional realization premises, every lowering uses the preservation contract of §10.3 with its side conditions discharged, and each used mode interface is sound and satisfies the stated coherence obligation. The evidence has a finite acyclic dependency graph complete for both versions. Let \(\Delta\) include every changed premise, rule instance, interface, interpretation and artifact binding, and let \(A\) be its forward closure through the old and revised dependencies.

Then the realized target of \(K[f']\) satisfies the transported safety judgment. Evidence records outside \(A\) retain their validity. Re-establishing affected records and recomposing the root yields evidence bound to the revised target.

\emph{Proof.} Proposition 7 gives observation inclusion for \(K[f']\) in \(K[f]\). Applying \(\pi\) to that inclusion establishes the revised source judgment. Induction along the realization path uses each lowering\textquotesingle s preservation contract and each admitted mode transport. Coherence relates the resulting interpreted claims where the compared paths meet. For evidence reuse, proceed in revised dependency order. A record outside \(A\) has unchanged direct supports and no affected predecessor, since either change would place it in \(A\). Its existing rule instance therefore remains valid. Re-established affected records compose with these retained records to certify the new target. Root or client certificates naming the previous artifact are rebound or reconstructed, using the unchanged generic rules and subproofs where applicable. \(\square\)

This proposition combines our component, transport and dependency interfaces. The compiler must establish each particular comparison and its use-site premises. A channel proof may support the comparison, while a buffer bound supplies a realization premise and an invocation identity connects the two. Progress and quantitative guarantees retain the additional contracts described in §§5 and 9. Keeping these premises with the affected components makes the proof\textquotesingle s scope visible at the place a developer changes the program.

Developer velocity follows from this division of work. A library\textquotesingle s general proof can serve many component instances, and each instance supplies its actual parameters and premises. A later edit revisits the affected applications and lowering scopes. The compiler accounts for changes in target declarations, theorem bindings and shared resources as well as source expressions. Stable evidence outside that dependency closure remains reusable, reducing repeated checking and compilation while preserving the reason each fragment was admitted.

For a live service, this includes the frames already suspended under an earlier revision. Their return points and storage remain associated with that revision until they complete, or an explicit migration establishes the new relation. New invocations can use a replacement implementation whose boundary contract has been admitted.

This gives formal information an efficiency role throughout the development cycle. Per-cut liveness reduces retained frame data. Established ranges justify representations. Applicable protocol laws avoid repeated per-project arguments. Dependency-directed revalidation limits repeated proof construction and lowering to the affected scope. These mechanisms apply the negative cost of abstraction reasoning discussed in our companion papers and by Stroustrup \hyperref[ref-stroustrup]{{[}31{]}}. Richer retained structure gives the compiler information with which to remove work.

\section*{11. Relationship to Established Work}\label{relationship-to-established-work}
\addcontentsline{toc}{section}{11. Relationship to Established Work}

The construction combines three traditions with distinct proof responsibilities. Session calculi describe communication structure, delimited control describes the remainder of a computation, and separation logic describes the resources those interactions use. Their connection is most useful when each imported result retains its own premises.

\subsection*{11.1 Stateful Sessions and Progress}\label{stateful-sessions-and-progress}
\addcontentsline{toc}{subsection}{11.1 Stateful Sessions and Progress}

Our service construction uses the coexponential client-pool interpretation with the M-True and QueW premises stated in §2.1. Under these rules, the authors\textquotesingle{} separation lemma represents independently typed components as parallel processes \hyperref[ref-qkb-errata]{{[}8{]}}. This supplies a premise of the progress argument for their client-server calculus, CSLL. We use the published preservation and progress results, Theorems 3.4 and 3.5, with this rule discipline \hyperref[ref-qkb]{{[}2{]}}. Their progress theorem uses the calculus\textquotesingle s reduction relation, including commuting conversions and reductions under prefixes. The actor transition correspondence in §4 therefore provides a separate operational obligation for a conventional scheduler.

The same care applies to the logical comparison. The paper\textquotesingle s Theorem 2.2 establishes coincidence of exponentials and coexponentials up to provability under both Mix and BiCut. The following discussion extends the comparison to the strong modalities. Both hypotheses matter. The authors also describe a terminating, deadlock-free extension of CP with Mix and Mix\(_0\), so the presence of Mix alone cannot determine that operational property. Our construction uses the distinct service and pool rules. An extension beyond that construction must establish its own preservation and progress properties under its chosen operational semantics.

Padovani supplies a subsequent fair-termination result for \(\mathrm{CSLL}^{\infty}\) \hyperref[ref-padovani-fair]{{[}32{]}}. Theorem 16 and Corollary 17 apply to validly typed processes with interface \(x:1\). Validity requires every infinite typing branch to revisit a server rule on the same channel infinitely often. Theorem 3 characterizes fair termination by the existence of a terminating continuation from every reachable process. The fairness notion used in that result concerns reduction runs. Applying it to a runtime requires an argument connecting that condition to the chosen scheduling policy.

This answers the termination question for Padovani\textquotesingle s calculus and typing discipline. Its syntax omits original CSLL constructs, including sequential client pools and traditional-exponential replicated servers, and its reduction semantics differs. The result therefore supplies an established termination theory against which to check an applicable service fragment. Using it for Clef requires a translation into valid derivations and an operational correspondence preserving its completion conditions. Our finite actor construction and the finite-batch completion lemma in §5.3 establish their conclusions under their own premises. For an indefinitely available service, the application requires completion of admitted requests while the service remains available for later work.

Our comparison with executable language designs includes Par. Its documentation explicitly relates the derived \lstinline!Cobang! and \lstinline!Coquest! types to Qian, Kavvos and Birkedal\textquotesingle s coexponentials, using its fan construction and native \lstinline!poll!/\lstinline!submit! control \hyperref[ref-par]{{[}33{]}}. A poll selects a ready client. Its active branch submits permitted descendants of that client and continues polling. This provides a concrete language precedent for expressing client-pool behavior through derived types and control primitives. Our construction connects a specified actor fragment to continuation frames and retains the resulting ownership, protocol and dependency facts through native lowering.

Our proof-library account has a corresponding distinction. Actris 2.0 establishes safety and dependent protocol reasoning for its channels \hyperref[ref-actris]{{[}17{]}}. LinearActris establishes global progress for its own channel calculus through a linear-resource and connectivity discipline \hyperref[ref-linearactris]{{[}34{]}}. This excludes a globally stuck unfinished configuration under its semantics. Completion of each admitted request additionally uses the premises in §5.3. Aneris and the reliable channel development supply the distributed interpretation needed when the same application crosses machine boundaries \hyperref[ref-aneris]{{[}18{]}} \hyperref[ref-rclib]{{[}19{]}}.

\subsection*{11.2 Reconstruction and Distributed Verification}\label{reconstruction-and-distributed-verification}
\addcontentsline{toc}{subsection}{11.2 Reconstruction and Distributed Verification}

Our finite communication summary addresses a defined extraction problem. Existing session inference work supplies stronger results for its own source languages and restrictions \hyperref[ref-spaccasassi]{{[}12{]}} \hyperref[ref-padovani-inference]{{[}13{]}}. Deniélou and Yoshida\textquotesingle s work connects compatible communicating automata to synthesis of global session types \hyperref[ref-denielou]{{[}35{]}}. These are useful sources for extending the compiler\textquotesingle s repertoire while retaining a precise account of the fragment each procedure covers.

Our contribution places such procedures alongside the value, lifetime and representation analyses already needed for native compilation. A discovered session shape can share its participants with the wait relation and captured-frame description. The subsequent proof uses those actual relationships, so an endpoint inferred at one stage remains identifiable at the operation that sends or receives through it.

We retain the verification motivation developed in the Erlang tradition. McErlang explores process behavior under an explicit execution environment and property specification \hyperref[ref-mcerlang]{{[}36{]}}. Compositional proofs in Iris and Aneris supply a complementary form of evidence. A parameterized channel or protocol theorem can cover every instance satisfying its premises, while execution exploration can investigate a particular composition and reveal missing assumptions. Our compiler account connects reusable theorem instances to the source operations and native realizations that consume them.

\subsection*{11.3 Computation, Control and Logical Duality}\label{computation-control-and-logical-duality}
\addcontentsline{toc}{subsection}{11.3 Computation, Control and Logical Duality}

Our continuation representation concerns the lifetime and resumption of a particular interaction. In the function/cofunction account of Choudhury and Gay, categorical coexponentials are binary constructions dual to function types, with coabstraction and coapplication given an operational interpretation \hyperref[ref-cofunction]{{[}3{]}}. Their accompanying artifact supplies executable implementations. In this paper, the continuation proof establishes that an invocation retains its identity, state and resources across suspension and resumption. The session modality and the continuation realization each retain their own typing and operational obligations.

Our negative and fractional types paper gives computational interpretations of directed recovery and value-specific resource discharge \hyperref[ref-nft]{{[}37{]}}. Those operations can occur within an actor handler and contribute their own resource and value laws. Their interpretation remains distinct from the coexponential service protocol. A continuation retaining such a value must preserve its admitted uses just as it preserves any other owned resource.

Our \emph{Adaptive Domain Models} paper supplies a further application in which stateful services perform adaptive computation \hyperref[ref-adm]{{[}21{]}}. In that setting, a changed algorithm, arithmetic regime or sampling policy has effects on both the result contract and the state carried between requests. The construction in this paper provides the communication and suspension account against which those changes can be checked.

The information-geometric pullback of Karwowski and colleagues supplies the metric used by that application \hyperref[ref-karwowski-2026]{{[}23{]}}. Its contravariance does not identify it with the session coexponential or with an executable inverse. Our connection is through an admitted numerical operation whose sampling, ownership and preservation obligations can be carried by the service.

\section*{12. Stateful Services in a Concurrent Language}\label{stateful-services-in-a-concurrent-language}
\addcontentsline{toc}{section}{12. Stateful Services in a Concurrent Language}

The coexponential construction makes a stateful service and its client pool available to type-theoretic reasoning. Our finite translation supplies the internal state protocol and worker derivation explicitly. The continuation account then identifies the state, resources and reply authority that must survive each suspension. Wait analysis, scheduling assumptions and application-specific laws supply the further conditions under which those conversations complete and their orders can change.

The distributed proof infrastructure makes this account useful beyond a local execution. Resource assertions travel through admitted protocol steps, indexed acceptance prevents duplicated traffic from duplicating ownership, and BAREWire connects each accepted message to its received meaning. Exact representation laws compose across those boundaries. Quantitative worker contracts add the accuracy of the returned computation. In the solver example, Urschel\textquotesingle s growth bound contributes a mathematical premise to the same request whose identity and resources are preserved across the network. The ADM service adds jointly sampled metric contributions and a restricted update, showing why a valid distributed protocol must retain the statistical construction as well as the returned values.

The categorical account connects these results to the framework\textquotesingle s broader construction. Parameterized composition retains the relationships within a computation. Our realization two-category composes the comparisons that justify replacing its components. Mode interfaces and compatible stage evidence keep those comparisons usable as the compiler changes the representation. A source request, its suspended continuation and its native completion can therefore share an inspectable chain of justification.

Composer uses this structure to select storage, preserve native execution and reuse unaffected work after an edit. Its engineering role extends from device-bound execution on a single core to vector, GPU, tiled and spatial computation, and to services distributed across machines. \emph{Fixed-Point Scaffolding} develops the passage from established facts to unboxed operations and differential compilation \hyperref[ref-fps]{{[}7{]}}. Here the stateful service gives those mechanisms a concrete concurrent application.

Our \emph{Decidable by Construction} paper describes the resulting engineering freedom as \emph{truly fearless} \hyperref[ref-dbc]{{[}6{]}}. A developer can change a service, specialize its worker or select another execution target while the compiler checks the contracts affected by that choice and retains the work whose justification survives. The coexponential account makes that freedom specific to concurrent software. Its protocol, continuation and representation laws supply reusable proof material, and their composition supports both the integrity of the resulting program and the pace at which it can be developed.

\begin{center}\rule{0.5\linewidth}{0.5pt}\end{center}

\section*{Software Availability}\label{software-availability}
\addcontentsline{toc}{section}{Software Availability}

The Clef language, Composer compiler, and supporting libraries described in this paper are developed under the Fidelity Framework project. Source repositories are available at \url{https://github.com/FidelityFramework}. The language specification, design rationale, and compiler documentation are published at \url{https://clef-lang.com}. Central components of the framework are dual-licensed; terms are detailed in each repository. The compiler design connects these components through the obligation and preservation interfaces described in this paper.

The propositions in this paper are manuscript proofs over the stated models and contracts. A Rocq-checked correspondence from Clef execution to the imported session and separation-logic models remains part of the proposed realization. The native bindings must also establish the operation, scheduling and delivery premises on which each application relies.

\section*{References}\label{references}
\addcontentsline{toc}{section}{References}

\leavevmode\protect\phantomsection\label{ref-armstrong}{}{[}1{]} Armstrong, J. (2003). \emph{Making Reliable Distributed Systems in the Presence of Software Errors}. PhD thesis, Royal Institute of Technology, Stockholm. \href{https://erlang.org/download/armstrong_thesis_2003.pdf}{Author\textquotesingle s thesis}.

\leavevmode\protect\phantomsection\label{ref-qkb}{}{[}2{]} Qian, Z., Kavvos, G. A., and Birkedal, L. (2021). Client-server sessions in linear logic. \emph{Proceedings of the ACM on Programming Languages}, 5(ICFP), Article 62. \href{https://doi.org/10.1145/3473567}{DOI:10.1145/3473567}. \href{https://cpv.au.dk/pdfs/2021-icfp-coexp-final.pdf}{Published paper}.

\leavevmode\protect\phantomsection\label{ref-cofunction}{}{[}3{]} Choudhury, V., and Gay, S. J. (2025). The duality of \(\lambda\)-abstraction. \emph{Proceedings of the ACM on Programming Languages}, 9(POPL), Article 12. \href{https://doi.org/10.1145/3704848}{DOI:10.1145/3704848}. \href{https://eprints.gla.ac.uk/342910/1/342910.pdf}{Published paper}. \href{https://github.com/vikraman/popl25-duality-artifact}{Accompanying artifact}.

\leavevmode\protect\phantomsection\label{ref-dts}{}{[}4{]} Haynes, H. (2026). Dimensional type systems and deterministic memory management: Design-time semantic preservation in native compilation. \href{https://arxiv.org/abs/2603.16437}{arXiv:2603.16437}.

\leavevmode\protect\phantomsection\label{ref-phg}{}{[}5{]} Haynes, H. (2026). The program hypergraph: Multi-way relational structure for geometric algebra, spatial compute, and physics-aware compilation. \href{https://arxiv.org/abs/2603.17627}{arXiv:2603.17627}.

\leavevmode\protect\phantomsection\label{ref-dbc}{}{[}6{]} Haynes, H. (2026). Decidable by construction: Design-time verification for truly fearless systems. \href{https://arxiv.org/abs/2603.25414}{arXiv:2603.25414}.

\leavevmode\protect\phantomsection\label{ref-fps}{}{[}7{]} Haynes, H. (2026). Fixed-point scaffolding in the Clef programming language: Our theoretical grounding for type-preserving compilation and proof inference. \href{https://arxiv.org/abs/2606.02854}{arXiv:2606.02854}.

\leavevmode\protect\phantomsection\label{ref-qkb-errata}{}{[}8{]} Qian, Z., Kavvos, G. A., and Birkedal, L. (2025, January 15). \emph{Errata for {\textquotedblleft{}Client-Server} Sessions in Linear {Logic\textquotedblright{}}}. \href{https://cs.au.dk/~birke/papers/coexp-conf-errata.pdf}{Authors\textquotesingle{} correction}.

\leavevmode\protect\phantomsection\label{ref-cdl}{}{[}9{]} Gavranović, B., Lessard, P., Dudzik, A. J., von Glehn, T., Araújo, J. G. M., and Veličković, P. (2024). Position: Categorical deep learning is an algebraic theory of all architectures. In \emph{Proceedings of the 41st International Conference on Machine Learning}, PMLR 235, 15209--15241. \href{https://proceedings.mlr.press/v235/gavranovic24a.html}{Publisher and paper}.

\leavevmode\protect\phantomsection\label{ref-danvy-filinski}{}{[}10{]} Danvy, O., and Filinski, A. (1990). Abstracting control. In \emph{Proceedings of the ACM Conference on Lisp and Functional Programming}, 151--160. \href{https://doi.org/10.1145/91556.91622}{DOI:10.1145/91556.91622}. \href{https://hjemmesider.diku.dk/~andrzej/papers/}{Author\textquotesingle s publication archive}.

\leavevmode\protect\phantomsection\label{ref-ocaml-effects}{}{[}11{]} Sivaramakrishnan, K. C., Dolan, S., White, L., Kelly, T., Jaffer, S., and Madhavapeddy, A. (2021). Retrofitting effect handlers onto OCaml. In \emph{PLDI}, 206--221. \href{https://doi.org/10.1145/3453483.3454039}{DOI:10.1145/3453483.3454039}. \href{https://arxiv.org/abs/2104.00250}{arXiv:2104.00250}.

\leavevmode\protect\phantomsection\label{ref-spaccasassi}{}{[}12{]} Spaccasassi, C., and Koutavas, V. (2016). Type-based analysis for session inference (extended abstract). In \emph{FORTE}, LNCS 9688, 248--266. \href{https://doi.org/10.1007/978-3-319-39570-8_17}{DOI:10.1007/978-3-319-39570-8\_17}. \href{https://www.scss.tcd.ie/Vasileios.Koutavas/publications/forte16.pdf}{Author manuscript}.

\leavevmode\protect\phantomsection\label{ref-padovani-inference}{}{[}13{]} Padovani, L. (2015). Type reconstruction for the linear \(\pi\)-calculus with composite regular types. \emph{Logical Methods in Computer Science}, 11(4), Article 13. \href{https://lmcs.episciences.org/1614}{Publisher and paper}.

\leavevmode\protect\phantomsection\label{ref-commutativity}{}{[}14{]} Koskinen, E., and Bansal, K. (2020). Reducing commutativity verification to reachability with differencing abstractions. \href{https://arxiv.org/abs/2004.08450}{arXiv:2004.08450}.

\leavevmode\protect\phantomsection\label{ref-dilworth}{}{[}15{]} Dilworth, R. P. (1950). A decomposition theorem for partially ordered sets. \emph{Annals of Mathematics}, 51(1), 161--166. \href{https://doi.org/10.2307/1969503}{DOI:10.2307/1969503}.

\leavevmode\protect\phantomsection\label{ref-iris}{}{[}16{]} Jung, R., Krebbers, R., Jourdan, J.-H., Bizjak, A., Birkedal, L., and Dreyer, D. (2018). Iris from the ground up: A modular foundation for higher-order concurrent separation logic. \emph{Journal of Functional Programming}, 28, e20. \href{https://www.mpi-sws.org/~dreyer/papers/iris-ground-up/paper.pdf}{Paper}.

\leavevmode\protect\phantomsection\label{ref-actris}{}{[}17{]} Hinrichsen, J. K., Bengtson, J., and Krebbers, R. (2022). Actris 2.0: Asynchronous session-type based reasoning in separation logic. \emph{Logical Methods in Computer Science}, 18(2), Article 16. \href{https://iris-project.org/pdfs/2022-lmcs-actris2-final.pdf}{Published paper}. \href{https://iris-project.org/actris/}{Development}.

\leavevmode\protect\phantomsection\label{ref-aneris}{}{[}18{]} Krogh-Jespersen, M., Timany, A., Ohlenbusch, M. E., Gregersen, S. O., and Birkedal, L. (2020). Aneris: A mechanised logic for modular reasoning about distributed systems. In \emph{ESOP}, LNCS 12075, 336--365. \href{https://doi.org/10.1007/978-3-030-44914-8_13}{DOI:10.1007/978-3-030-44914-8\_13}. \href{https://iris-project.org/pdfs/2020-esop-aneris-final.pdf}{Paper}.

\leavevmode\protect\phantomsection\label{ref-rclib}{}{[}19{]} Gondelman, L., Hinrichsen, J. K., Pereira, M., Timany, A., and Birkedal, L. (2023). Verifying reliable network components in a distributed separation logic with dependent separation protocols. \emph{Proceedings of the ACM on Programming Languages}, 7(ICFP), Article 217. \href{https://doi.org/10.1145/3607859}{DOI:10.1145/3607859}. \href{https://iris-project.org/pdfs/2023-icfp-relcom.pdf}{Paper}.

\leavevmode\protect\phantomsection\label{ref-tarau}{}{[}20{]} Tarau, P. (2009). \emph{Isomorphic Data Encodings in Haskell and their Generalization to Hylomorphisms on Hereditarily Finite Data Types}. \href{https://arxiv.org/pdf/0808.2953v4}{arXiv:0808.2953v4}. First posted 2008; revised January 19, 2009.

\leavevmode\protect\phantomsection\label{ref-adm}{}{[}21{]} Haynes, H. (2026). Adaptive domain models: Bayesian evolution, warm rotation, and principled training for geometric and neuromorphic AI. \href{https://arxiv.org/abs/2603.18104}{arXiv:2603.18104}.

\leavevmode\protect\phantomsection\label{ref-urschel}{}{[}22{]} Urschel, J. (2026). On the growth factor of random matrices. \href{https://arxiv.org/abs/2610.06785v1}{arXiv:2610.06785v1}.

\leavevmode\protect\phantomsection\label{ref-karwowski-2026}{}{[}23{]} Karwowski, J., Nielsen, F., and Staton, S. (2026). Compositional information geometry of probabilistic programs. \href{https://arxiv.org/abs/2610.17886v1}{arXiv:2610.17886v1}.

\leavevmode\protect\phantomsection\label{ref-phase}{}{[}24{]} Harper, R., Mitchell, J. C., and Moggi, E. (1990). Higher-order modules and the phase distinction. In \emph{POPL}, 341--354. \href{https://doi.org/10.1145/96709.96744}{DOI:10.1145/96709.96744}. \href{https://theory.stanford.edu/~jcm/papers/harper-mm-90.pdf}{Author manuscript}.

\leavevmode\protect\phantomsection\label{ref-havarneanu}{}{[}25{]} Hăvărneanu, A. (2026). \emph{Classical Adjoint Logic}. Pre-publication manuscript, July 12.

\leavevmode\protect\phantomsection\label{ref-lack}{}{[}26{]} Lack, S. (2007). \emph{A 2-categories companion}. \href{https://arxiv.org/abs/math/0702535v1}{arXiv:math/0702535v1}.

\leavevmode\protect\phantomsection\label{ref-sheaf}{}{[}27{]} Ayzenberg, A., Gebhart, T., Magai, G., and Solomadin, G. (2025). \emph{Sheaf theory: from deep geometry to deep learning}. \href{https://arxiv.org/abs/2502.15476v1}{arXiv:2502.15476v1}.

\leavevmode\protect\phantomsection\label{ref-gpu-execution}{}{[}28{]} NVIDIA. \emph{CUDA Programming Guide}. Sections 2.3, 2.5 and 4.3.2.2, accessed October 8, 2026. \href{https://docs.nvidia.com/cuda/cuda-programming-guide/02-basics/writing-cuda-kernels.html}{SIMT execution}. \href{https://docs.nvidia.com/cuda/cuda-programming-guide/02-basics/asynchronous-execution.html}{Asynchronous execution}. \href{https://docs.nvidia.com/cuda/cuda-programming-guide/04-special-topics/stream-ordered-memory-allocation.html\#freeing-memory}{Allocation lifetime}.

\leavevmode\protect\phantomsection\label{ref-aie}{}{[}29{]} AMD. (2026, February 18). \emph{Versal Adaptive SoC AI Engine Architecture Manual (AM009)}, revision 1.4. \href{https://docs.amd.com/r/en-US/am009-versal-ai-engine/AI-Engine-Tile-Architecture}{AI Engine Tile Architecture}.

\leavevmode\protect\phantomsection\label{ref-islaris}{}{[}30{]} Sammler, M., Hammond, A., Lepigre, R., Campbell, B., Pichon-Pharabod, J., Dreyer, D., Garg, D., and Sewell, P. (2022). Islaris: Verification of machine code against authoritative ISA semantics. In \emph{PLDI}. \href{https://doi.org/10.1145/3519939.3523434}{DOI:10.1145/3519939.3523434}. \href{https://www.cl.cam.ac.uk/~pes20/2022-pldi-islaris.pdf}{Published paper}.

\leavevmode\protect\phantomsection\label{ref-stroustrup}{}{[}31{]} Peterman, R. (2026, May 18). \href{https://www.developing.dev/p/creator-of-c-bell-labs-negative-overhead}{Creator of C++: Bell Labs, Negative Overhead Abstraction, Mistakes {|} Bjarne Stroustrup}. Interview with Bjarne Stroustrup. \emph{The Peterman Pod}. Transcript, 1:22:18--1:23:42.

\leavevmode\protect\phantomsection\label{ref-padovani-fair}{}{[}32{]} Padovani, L. (2023). On the fair termination of client-server sessions. In \emph{28th International Conference on Types for Proofs and Programs (TYPES 2022)}, LIPIcs 269, 5:1--5:21. \href{https://doi.org/10.4230/LIPIcs.TYPES.2022.5}{DOI:10.4230/LIPIcs.TYPES.2022.5}. \href{https://drops.dagstuhl.de/storage/00lipics/lipics-vol269-types2022/LIPIcs.TYPES.2022.5/LIPIcs.TYPES.2022.5.pdf}{Published paper}. \href{https://arxiv.org/abs/2212.05457}{2022 preprint}.

\leavevmode\protect\phantomsection\label{ref-par}{}{[}33{]} Par contributors. \emph{Par language documentation: Poll and Submit; The Fan Pattern}. Revision 30decb7f9e59, accessed October 8, 2026. \href{https://github.com/par-team/par-lang/blob/30decb7f9e597e3384e383623b0e82a89a88956d/docs/src/nondeterminism/poll_submit.md}{Polling semantics}. \href{https://github.com/par-team/par-lang/blob/30decb7f9e597e3384e383623b0e82a89a88956d/docs/src/nondeterminism/fan_pattern.md}{Coexponential construction}.

\leavevmode\protect\phantomsection\label{ref-linearactris}{}{[}34{]} Jacobs, J., Hinrichsen, J. K., and Krebbers, R. (2024). Deadlock-free separation logic: Linearity yields progress for dependent higher-order message passing. \emph{Proceedings of the ACM on Programming Languages}, 8(POPL), Article 47. \href{https://doi.org/10.1145/3632889}{DOI:10.1145/3632889}. \href{https://iris-project.org/pdfs/2024-popl-dlfactris.pdf}{Paper}.

\leavevmode\protect\phantomsection\label{ref-denielou}{}{[}35{]} Deniélou, P.-M., and Yoshida, N. (2013). Multiparty compatibility in communicating automata: Characterisation and synthesis of global session types. In \emph{ICALP}, LNCS 7966, 174--186. \href{https://arxiv.org/abs/1304.1902}{arXiv:1304.1902}.

\leavevmode\protect\phantomsection\label{ref-mcerlang}{}{[}36{]} Fredlund, L.-Å., and Svensson, H. (2007). McErlang: A model checker for a distributed functional programming language. In \emph{ICFP}, 125--136. \href{https://doi.org/10.1145/1291151.1291171}{DOI:10.1145/1291151.1291171}.

\leavevmode\protect\phantomsection\label{ref-nft}{}{[}37{]} Haynes, H. (2026). Negative and fractional types in the Fidelity Framework: Computation, reversibility and resource pairing. \href{https://arxiv.org/abs/2606.04352}{arXiv:2606.04352}.
\end{document}